\documentclass[runningheads]{llncs}

\usepackage{blindtext} 
\usepackage{booktabs} 
\usepackage[shortlabels]{enumitem}
\usepackage{graphicx}
\usepackage[caption=false,font=scriptsize,labelfont=sf,textfont=sf]{subfig}

\usepackage{mathtools}
\usepackage{tcolorbox}
\usepackage{enumitem}

\usepackage{comment}

\setlist[itemize]{noitemsep} 

\usepackage[colorlinks,pagebackref,breaklinks]{hyperref}
\hypersetup{
  hypertexnames=true, linkcolor=blue, anchorcolor=black,
  citecolor=blue, urlcolor=blue  
}
\usepackage{balance}

\newcommand{\BOT}{\texttt{BOT}}
\newcommand{\SHORT}{\texttt{MetaPriv}}

\begin{document}

\title{MetaPriv: Acting in Favor of Privacy on Social Media Platforms}

\author{Robert Cantaragiu \and Antonis Michalas \and Eugene Frimpong \and \\Alexandros Bakas}
\authorrunning{R. Cantaragiu et al.}
%
\institute{Tampere University, Tampere, Finland\\
\email{\{robert.cantaragiu, antonios.michalas, eugene.frimpong, alexandros.bakas\}@tuni.fi}}

\maketitle

\begin{abstract}
Social networks such as Facebook\footnote{Since October 2021 is also known as META.} (FB) and Instagram are known for tracking user online behaviour for commercial gain. To this day, there is practically no other way of achieving privacy in said platforms other than renouncing their use. However, many users are reluctant in doing so because of convenience or social and professional reasons. In this work, we propose a means of balancing convenience and privacy on FB through obfuscation. We have created \SHORT{}, a tool based on simulating user interaction with FB. \SHORT{} allows users to add noise interactions to their account so as to lead FB's profiling algorithms astray, and make them draw inaccurate profiles in relation to their interests and habits. To prove our tool's effectiveness, we ran extensive experiments on a dummy account and two existing user accounts
. Our results showed that, by using our tool, users can achieve a higher degree of privacy in just a couple of weeks
. We believe that \SHORT{} can be further developed to accommodate other social media platforms and help users regain their privacy, while maintaining a reasonable level of convenience. To support open science and reproducible research, our source code is publicly available online.
\end{abstract}


\section{Introduction}
\label{sec:Intro}
Online tracking on social networks (SNs) have raised concerns regarding user privacy~\cite{onavo,bug,cambridge}. Recommendation systems used by social media 
are developed to present biased information with the purpose of encouraging user engagement. When users share their opinions, beliefs and preferences on said platforms -- whether by clicking `like' on an article or by writing a controversial post -- the recommendations they receive are aimed at reinforcing these beliefs. Their goal is to provide users with information that most likely interests them and enables them to trace other users sharing the same values. It is believed that through this approach, users gradually become more engaged with these platforms, while 
going deeper in the rabbit-hole of subjectivity, since the only information and news they receive affirms their already established opinions. As a result, users remain engaged in SN platforms, as the latter  make accurate predictions on their potential consumption needs. Hence, platforms in collaboration with companies promoting their products manipulate user information for targeted advertising.

\smallskip

\noindent\textit{\underline{Balance between Privacy and Convenience on Social Networks}:}
Most users seem to be left with two options when it comes to social network privacy: \textit{(1)} either regular use of the platform -- hence no privacy or \textit{(2)} complete abstinence from social networks -- hence full privacy. However, the second option presents a number of problems. First, the hassle of removing data about oneself from a platform, discourages users as it demands tedious action. Note that data removal does not refer to deleting the account alone, but to the deletion of all posts, pictures and logged data from the platform. Secondly, even in cases where all user data is deleted, SNs may still track individuals through partner companies on different websites (e.g.\ through FB Pixel~\cite{pixel}). Finally, completely opting out of SNs results in great costs in terms of convenience for many individuals, who wish to keep in touch with their friends, keep up with the news and promote themselves or their activities. To this end, we believe that complete privacy is not achievable for most users. We do, however, think that one can strike a balance between privacy and convenience on said platforms and this has been a major motive behind our work. Our platform of choice for this work is FB -- the world’s largest online SN. However, the idea presented below can be developed to accommodate privacy on other platforms. 

\smallskip

\noindent\textit{Contributions:}
The main idea 
has been developed based on increasing concerns regarding the breach of user privacy in online SNs. More precisely, the main concern is that user choices are being covertly manipulated and controlled by SNs. With this in mind, we built \SHORT, an automated tool that allows FB users to obfuscate their data and conceal their real interests and habits from FB. As a result, the core contribution of this paper is that it provides users with the necessary tools to protect their privacy when using SNs. It is worth mentioning that \SHORT{} allows users to define the desired level of privacy (e.g.\ become almost 'invisible' online while still using SN platforms, reveal certain information about their digital and real lives etc.). By doing this, \SHORT{} provides a novel and adaptive balance between privacy and functionality. This is a feature we believe will be used in several services in the near future. 


\section{Related work}
\label{sec:RelWork}



A number of research works offer users a more private experience on FB and other SNs.
\texttt{FaceCloak}~\cite{facecloak} protects user privacy on SNs by shielding personal information from the SN and unauthorized users, while maintaining the usability of the underlying services. \texttt{FaceCloak} achieves this through providing fake information to the SN and storing sensitive data in an encrypted form on a separate server. It is implemented as a Firefox 
extension for FB. \texttt{FaceCloak}'s user privacy attempt resembles our work. However, its main purpose is to hide specific data such as age, name, 
etc.\ and not user interests derived from interaction with the SN. Moreover, as of August 2011, the current version of the FaceCloak Firefox extension does not work with FB anymore due to changes made by FB~\cite{facecloak_down}. 

\texttt{Scramble}~\cite{scramble} allows users to enforce access control over their data. It is an SN-independent Firefox extension allowing users to define access control lists (ACL) of authorised users for each piece of data, based on their preferences. In addition to that, it also allows users to encrypt their posted content in the SN, therefore guaranteeing confidentiality of user data against the SN. The tool allows users to hide information through cryptography. This may require prior knowledge, which is usually counter intuitive for ordinary users. Also, it's implementation cannot be found anywhere and is likely outdated. 

Other privacy approaches focus on different platforms: Google, Youtube, Amazon etc. While they do not necessarily provide solutions for achieving privacy on FB, their approaches served as an inspiration for our work.

 \texttt{TrackThis}~\cite{trackthis} by Mozilla proposes an approach of polluting a users browsing history by opening 100 tabs at once. This leaves 
 cookies that are unrelated to the users interests and confuse third party trackers. Similarly, the authors of~\cite{182952} and~\cite{xing2013take} show a way to attack personalization algorithms by polluting a users browser history with noise by generating false clicks through cross-site request forgery (XSRF). In~\cite{kim2018adbudgetkiller}, the authors present an attack for draining ad budgets. By repeatedly pulling ads using crafted browsing profiles, they managed to  reduce the chance of showing their ads to real
 visitors and trash the ad budget. While having similar approaches to ours, these tools provide limited privacy in the long run as they have to be relaunched after a period of time.
 
  In~\cite{data_strikes}, the authors test protesting against data labouring~\cite{data_labor}: they utilize user interactions with different services as input for training user profiling algorithms. They simulate data strikes against recommendation systems. 
  Their results imply that data strikes can put a certain pressure on technology companies and that users have more control over their relationship with said companies. Our work can also be viewed as a protest against the data labouring of users on an SN: if enough users had access to noise attributes, the recommendation systems of FB would most likely be disrupted even for new users not using our tool.

 Howe and Nissenbaum proposed 
 \texttt{AdNauseam}~\cite{AdNauseam} -- a 
 browser extension designed to obfuscate browsing data and protect user-tracking by advertising networks. It clicks on every displayed ad in different web pages, thereby diminishing the value of all ad clicks -- obfuscating the real with fake clicks.
 Another tool called \texttt{Harpo}~\cite{zhang2021harpo} uses reinforcement learning to adaptively interleave real page visits with fake pages to distort a tracker's view of a user's browsing profile. \texttt{Harpo} is also able to achieve better stealthiness to adversarial detection as compared to \texttt{AdNauseam}. Our tool is designed and based on similar obfuscation ideas, however we focus on a specific SN platform and not only on advertisements. 
 

\section{System Model}
\label{sec:SysMod}
We now proceed with introducing the system model we consider by describing the main entities participating in the design of \SHORT, as well as their capacities. 

\smallskip

\noindent\textbf{Social Network (SN):} Defined as a graph $\mathcal{G} = (\mathcal{U}, \mathcal{R})$ where the vertices are comprised of users from a set $\mathcal{U}$, with the edges being the relationship between said users, described by the set $\mathcal{R} \subseteq \{ \{ u,v \} \,|\, u,v \in \mathcal{U} \, \textrm{and} \, u \neq v \} $.

\smallskip

\noindent\textbf{Users:} Let $\mathcal{U} := \{ u_1, \ldots, u_n \}$ be the set of all users registered in an online SN such as FB. Each user has a unique identifier $i \in [1,n]$. In addition to that, each user is associated with a number of attributes. The set of all attributes associated with a user $u_i$ is denoted as $\mathcal{A}_i \subseteq \mathcal{A}$. 

\smallskip

\noindent\textbf{Attributes:} The set of all available attributes in an SN is denoted by $\mathcal{A} := \{ a_1, \ldots, a_m\}$ and is called the attribute space. An attribute is a specific trait that a user $u_i$ possesses, e.g.\ ``$u_i$ likes cats".

\smallskip

\noindent\textbf{BOT:} An entity that adds noise to a user profile ($u_i$). It works by mimicking the user's interaction with the SN and generates noise attributes on their behalf.

\smallskip

\noindent\textbf{User Real and Noise Attributes:} Assume a user $u_i$ with a list of attributes $\mathcal{A}_i$. Elements of $\mathcal{A}_i$ may have been generated legitimately (i.e.\ through the user's real activity) or by the \BOT. The set of all attributes generated by the user's legitimate activity is denoted as $\mathcal{A}_i^r \subseteq \mathcal{A}_i$ while the set of all attributes associated with $u_i$ but generated by the \BOT{} is denoted by $\mathcal{A}_i^n \subseteq \mathcal{A}_i$

\subsection{High-Level Overview}
\label{sub:HighLevel}

The core idea behind \SHORT{} is to fuddle FB's opinion about a user $u_i$ by obfuscating $u_i$'s real attributes $\mathcal{A}_i^r$  with the help of noise attributes $\mathcal{A}_i^n$. To that end, we use the \BOT{} and have it interact with the SN on behalf of $u_i$. Ideally, to achieve privacy, the amount of traffic generated by the \BOT{} should be the same or more than the traffic generated by $u_i$. 

When user $u_i$ creates an account on FB, they have no attributes (i.e.\ the set $\mathcal{A}_i$ is empty). Following registration, $u_i$ begins generating activity (e.g.\ adding friends, liking pages and posts).  
By collecting and analyzing user activities, FB creates a list of attributes that represents each user's perceived interests (e.g., $a_1$ -- ``$u_i$ likes cooking").  
For the purposes of this work, we consider 
these attributes as real and are added to the set $\mathcal{A}_i^r$ -- a subset of $\mathcal{A}_i$, i.e. $\mathcal{A}_i^r \subseteq \mathcal{A}_i$. 
The set $\mathcal{A}_i$ is then used by FB to decide which posts and advertisements are presented in the respective $u_i$ feed. In this scenario, all $u_i$'s interests are known to the SN, which can make accurate predictions about their preferences and therefore populate their account with accurate personalized content. In this work, we are examining ways of protecting user privacy from a potentially malicious or at least curious SN. To achieve this, we have created \SHORT{}.
With our tool, users can confuse an SN about their real interests. \SHORT{} revolves around a simple idea: Since the SN personalizes users by analyzing their activities on the platform, our tool generates noise traffic on behalf of a user. This will result in adding attributes to the set $\mathcal{A}_i^n$ containing the noise attributes described earlier. With this in mind, we  built a \BOT{} as part of the core of \SHORT{} whose functionality is described below. At this point, it is worth noting that the interactions generated by \SHORT{} consists of primarily liking posts and pages.

\begin{figure}[h]
    \centering
    \includegraphics[width=0.6\textwidth]{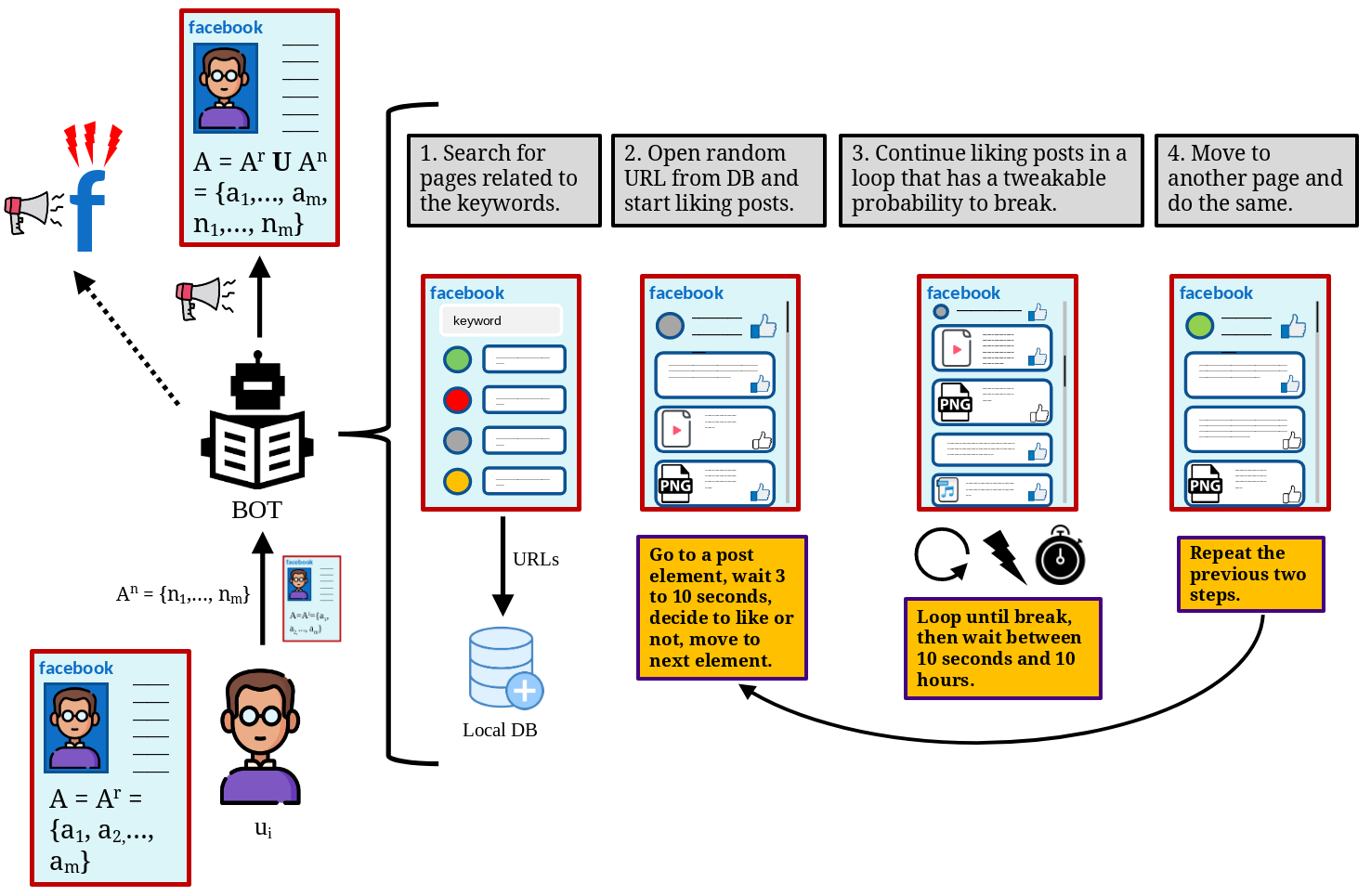}
    \caption{High-level overview of the \BOT{}'s functionality.}
    \label{fig:bot}
\end{figure}
\begin{enumerate}
    \item As a primary requirement, the \BOT{} needs access  to $u_i$'s account. This can be done in one of two ways: Either with $u_i$ providing their credentials 
    or through their browser profile folder i.e.\ the hidden folder in an operating system's user folder, where all web browser cookies,  etc.\ are stored. 
    
    \item Once the \BOT{} has gained access to the user account, it requires a set of keywords generated by a different part of \SHORT{}, which would serve as noise attributes. The keyword generator, however, requires a seed keyword that the user must input at least once.
    
    \item The user then inputs their desired level of privacy. This privacy level simply refers to the level of convenience and benefits that a user is willing to accommodate to better protect their privacy. In practice, it represents the amount of noise that is persistently added to an account.
    
    \item Finally, the \BOT{} repetitively executes a series of steps represented in \autoref{fig:bot}. 
\end{enumerate}

\subsection{Extending \SHORT{}}
\label{subsec:extension}
\begin{figure}[h]
    \centering
    \includegraphics[width=0.6\textwidth]{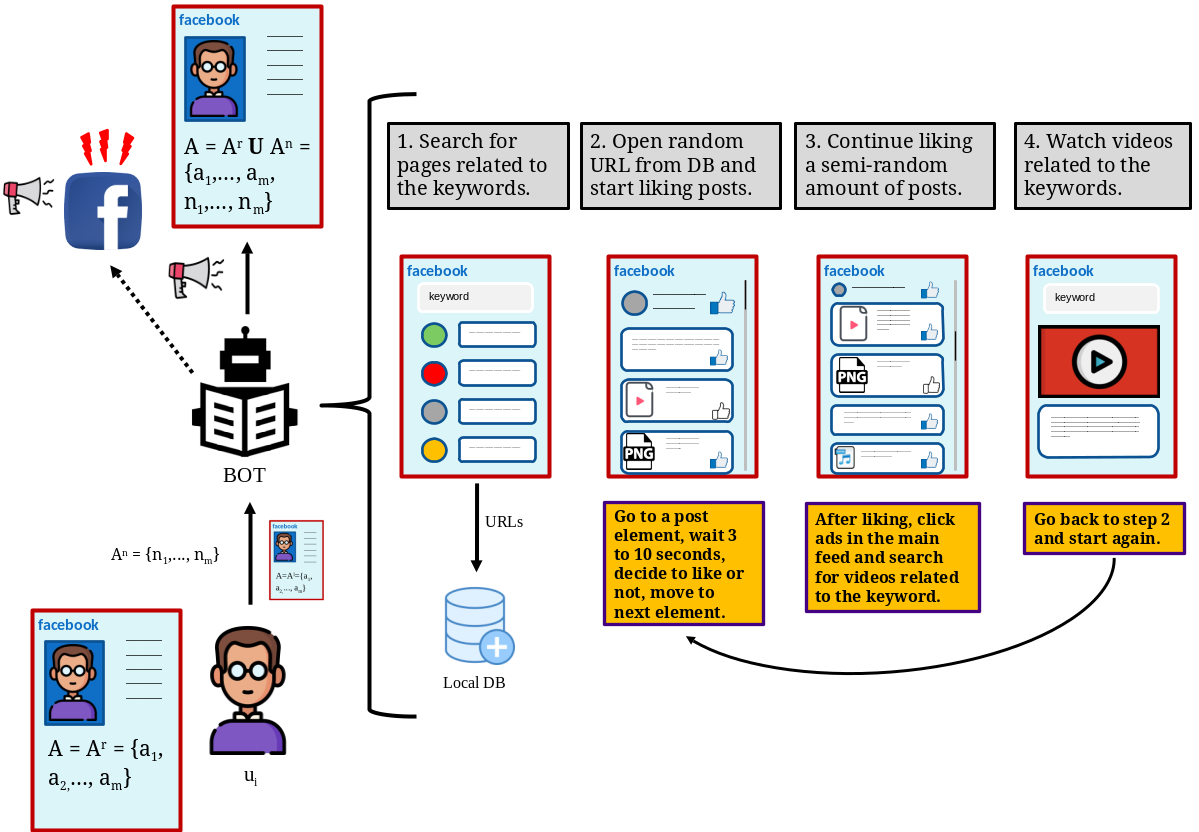}
    \caption{Extended \BOT{} Functionality.}
    \label{fig:extendbot}
\end{figure}

After extensive experiments, we observed limited success with the initial version of \SHORT{}, which we attributed to its limited interactions (i.e., simply liking posts and pages). These results are discussed in \autoref{sec:Results}. As such, it became necessary to add extra features to \SHORT. To limit the amount of noise generated, before the \BOT{} switches to another page, it waits for a random amount of time. In the basic implementation, \SHORT{} did not run any tasks during this wait period. However, in this extended version, \SHORT{} watches keyword related videos and clicks Facebook ads displayed in user's main feed instead of simply waiting. Our observations showed that video watching did not seem to raise any suspicions from FB, i.e.\ the browser session did not get logged out or blocked, hence the \BOT{} clicks on every ad from the first 100 posts in the main feed, searches the keyword in FB's video page 
and watches all the videos returned. This, we believe, helps to further reinforce the noise and give it more variety. \autoref{fig:extendbot} provides an overview of the extended functionalities of \SHORT{}.

\section{Measuring User Privacy on Facebook}
\label{sec:QuantifyPrivacy}

Previous works focus on measuring privacy according to the visibility and sensitivity of user attributes~\cite{aghasian2018users,maximilien2009enabling,domingo2010rational}. 
This approach, however, is inapplicable, as the aim is to confuse the data collector, thus leading to inaccurate user profile predictions. Visibility of a user's attributes would always be maximum, since the SN stores all user interactions with it. Additionally, in this work the concept of sensitivity cannot apply, since all user attributes are known to the SN (i.e.\ can be considered public). With this in mind, we propose a new definition for privacy on an SN based on a user's \textit{real} and \textit{noisy} interactions with the SN. Real interactions are daily, \textit{legitimate} user interactions with the SN. Noisy interactions are  \BOT{}-produced and mainly generate fake activity on a user's profile. 

Our first approach on quantifying privacy was characterized by rather elementary and naive thinking: Initially, we defined the notion of \textit{Theoretical Privacy}. The intuition behind Theoretical Privacy was that a user's level of privacy is proportional to the number of noise in their profile. However, the results of our first experiments did not support this. Apparently, the time that a user likes a post, a page, etc. seems to be significant for FB's personalization algorithms. More precisely, it seems that FB weighs a user's recent rather than older content. In view of the above, we refined our idea on quantifying privacy and defined \textit{Effective Privacy} -- an alternative that better fits FB's models.


\begin{definition}[Theoretical Privacy]\label{def:TPrivacy} Theoretical privacy is measured by taking into account the amount of posts liked by a user $u_i$ and the \BOT. 
User $u_i$'s theoretical privacy with $j + k$ attributes is defined as:

\begin{equation}
	P^{th}_i = \frac{\sum_{j \in \mathcal{A}_i^r} RA^{th}_{j} - \sum_{k \in \mathcal{A}_i^n} NA^{th}_{k}}{T}, 
\end{equation}

 \noindent where $RA_{th}$ is the number of specific attribute-related posts liked by $u_i$, $NA_{th}$ is the number of specific attribute-related posts liked by the \BOT{}  and $T$ is the total number of posts liked by $u_i$'s account.
 \end{definition}
 
 \begin{definition}[Effective Privacy]\label{def:EFPrivacy}
 For this definition we consider the effective strength of user real and noise attributes. The strength of a user's real attribute is proportional to:
 
\begin{itemize}
    \item the number of posts in the main feed from liked pages linked to an attribute. Variable: $r_p$
    \item the number of recommended, suggested and sponsored posts in the main feed from pages linked to an attribute, but not liked by the user or the \BOT . Variable: $r_{sp}$
    \item the number of video posts from the main video feed (\url{https://www.facebook.com/watch}) linked to an attribute. Variable: $r_{vp}$
    \item the number of video posts from the latest video feed (\url{https://www.facebook.com/watch/latest}) linked to an attribute. Variable: $r_{lvp}$
\end{itemize}

The effective strength of a real attribute is defined as:

\begin{equation}
	RA_{eff} = \frac{1}{n}\left(a\frac{r_p}{t_p}+b\frac{r_{sp}}{t_{sp}}+c\frac{r_{vp}}{t_{vp}}+d\frac{r_{lvp}}{t_{lvp}}\right),
\end{equation}

\noindent where  $a, b, c, d \in \{0,1\}, n = a+b+c+d$,  $t_p$ is the total number of posts shown in the main feed, $t_{sp}$ is the total number of suggested posts shown in the main feed, $t_{vp}$ is the total number of video posts related to $u_i$'s attributes from the main video feed  and $t_{lvp}$ is the total number of video posts from the latest video feed. Each of the variables $a, b, c, d$ is given the value 0, when their respective fraction is 0. Otherwise they are given the value 1. This is done so that, if one effective strength variable has a value of 0 (i.e. no posts), then it will not be taken into account for the final effective privacy value.

A similar definition stands for the effective strength of noise attributes $NA_{eff}$.  variables $r_p, r_{sp}, r_{vp}$ and $r_{lvp}$ are replaced with corresponding noise attributes i.e. $n_p, n_{sp}, n_{vp}$ and $n_{lvp}$. The strength of a noise attribute is defined as:

\begin{equation}
	NA_{eff} = \frac{1}{n}\left(a\frac{n_p}{t_p}+b\frac{n_{sp}}{t_{sp}}+c\frac{n_{vp}}{t_{vp}}+d\frac{n_{lvp}}{t_{lvp}}\right)
\end{equation}

Finally, for a user $u_i$ with $j + k$ attributes, we combine the two variables and reach the effective privacy:

\begin{equation}
	P^{eff}_i = \sum_{j \in \mathcal{A}_i^r} RA_{j}^{eff} - \sum_{k \in \mathcal{A}_i^n} NA_{k}^{eff}
\end{equation}
\end{definition}

In both cases, the resulting value will be $P \in [-1,1]$. The closer it is to 0, the more indistinguishable will the noise attributes be from real attributes. 
Therefore, the account of an arbitrary user $u_i$ is private \textit{iff} $P \approx 0$ or $P \leq 0$.

\section{Implementation and Results}
\label{sec:Results}

To demonstrate \SHORT's functionality and practicality, we evaluated both the basic and the extended versions. For the basic version, we created a dummy FB account and ran a~10-week experiment to build the account's real and noise attributes. While for the extended version, we tested \SHORT{} on two real FB accounts that have existed and are active for over a decade. 
To evaluate the dummy account, we used \SHORT{} to simulate both user and \BOT{} interactions\footnote{We make a clear distinction between \SHORT{} and the \BOT{}. \BOT{} interactions will be used to refer to the noise traffic generated by \SHORT{}.} with FB. Our test program was implemented using Python~3.10 and \href{https://www.selenium.dev/documentation/webdriver/}{Selenium WebDriver} -- a framework for testing web applications that allowed us to simulate an automated user interaction with FB. 

\smallskip

\noindent\textit{\textbf{Open Science and Reproducible Research:}}
Our source code\footnote{\url{https://github.com/ctrgrb/MetaPriv}} has been anonymized and made publicly available online to support open science and reproducible research. 


\subsection{Dummy Account Results}
\label{subsec:newaccount}

For the dummy FB account, we created a 22-year-old female user from Ireland (the account and all interactions were made through an Azure server with an Irish IP address). 
At the end of each week, we ran an extensive analysis of FB's main, video and latest video feed by opening the respective URLs, going through a certain amount of posts in them and saving the information about said posts in an SQL database. 

\noindent\textbf{Weeks 1 \& 2:} The first two weeks primarily consisted of building the user profile with a single attribute. To be more specific, we used the attribute ``cat", so FB would associate our user with cats. We then provided the keyword ``cat pictures" as input to \SHORT{}. The program liked 1,056 posts from 51 keyword-related pages over these two weeks. This keyword served as the user's \textit{real attribute}. 
After one week, 'Recommended' posts appeared in the main feed. Out of 264 posts, 32 were recommended and 11 seemed relevant to the user's profile:
\vspace{-0.5em}
\begin{quote}
     1 post related to demographics -a house in Dublin; 1 post about cats from a page about cats; 2 posts about tigers (both from FB group: WildCat Ridge Sanctuary); 1 post about demographics and cats (page name: North Dublin Cat Rescue Ireland); 1 post about ostriches, 1 about bulls, 2 about dogs, 1 about rare animals (related to animals); 1 post about ``Dads Acting Like Their Teenage Daughters" (possibly gender-related).
\end{quote}

\noindent Other recommended posts were unrelated to ``cats" and had a dozen million views (we assume these were most likely trending posts). Almost all the recommended posts were videos\footnote{This could be because users show a higher rate of engagement to online videos compared to text (e.g.\ articles, blog posts, etc.)}. After these two weeks, we analyzed 449 posts from the main feed and got 13 recommended posts along with 23 ``join group" recommendations from cat-related FB groups. 8 of the recommended posts were linked to the user's profile:
\vspace{-0.5em}
\begin{quote}
    1 post related to demographics: Football game GERMANY vs 
    IRELAND (2002); 1 post about cats from FB group: CAT LOVERS PHILIPPINES; 4 posts about animals from a group about animal comics; 1 post about cats from the 'Daily Mail' page; 1 post from a group about Dinosaurs. The name of the person posting was: Margaret Happycat.
\end{quote}

\noindent This time, most recommendations appeared from groups, though the user was not a member of any.

\noindent\textbf{Week 3:} For the third week, we added a second keyword as a noise attribute to the profile. At this point, the noise was manifested through liking a noise-related page and its posts at every 10th page switch. In essence, 10\% of the interactions with FB were now related to a single noise attribute. This 10\% represented 72 out of 554 posts liked in week~3 from~5 pages linked to the chosen noise keyword ``guns"\footnote{It is worth noting that the percentage value is an approximation since \SHORT{} is designed with randomness in mind to avoid patterns in its behaviour.}. We observed that there were no recommended posts after this period. An analysis of~547 posts from the main feed showed that 19 were linked to the noise attribute. The latest video feed contained only~21 videos from liked pages related to the real attribute (i.e.\ cats). In the main video feed, we analyzed~184 video posts.~70 of them included words such as: [`cat',`Cat',`kitten',`Kitten'] in their description or page URL and were, thus, related to the real attribute, while nothing was related to the noise attribute.

\noindent\textbf{Week 4:} For this period, we increased the noise amount from 10\% to 20\%. Out of 530 liked posts,~112 came from~8 pages related to the noise attribute. In the main feed, out of~337 posts,~38 were from pages related to the noise attribute. FB stopped showing recommended posts at this point, however, `Suggested for you' posts began to show. Out of the~337 posts,~8 were labeled as `Suggested' out of which~1 was related to animals,~3 specifically to cats and the remaining were possibly gender-related. This time too, the latest video feed showed only cat-related videos and in the main video feed, out of ~152 videos,~35 included the words: [`cat',`Cat',`kitten',`Kitten'] in the description or page URL, while no videos were related to guns.

\noindent\textbf{Week 5:} We decided to add another noise attribute, thus dividing FB interaction as follows: 70\% cats, 20\% guns and 10\% cooking. From a total of~485 liked posts,~130 were related to the keyword ``guns" and~36 to ``cooking recipes". This time, out of~673 posts in the main feed,~67 were related to guns and~147 to cooking. Our theory for increased cooking content is that a cat lover is more likely to also like cooking rather than guns\footnote{This might also be related to the fact that Ireland has one of Europe's least permissive firearm legislation  -- hence gun-related content is heavily regulated.}. 
This time, out of~16 suggested posts,~14 were cats. In the latest video feed, out of~51 videos,~21 were cats,~1 guns and~26 cooking. Finally, in the main video feed, out of~136 posts,~27 were cats,~3 guns and~7 cooking.

\noindent\textbf{Week 6:} We increased the amount of noise for the cooking attribute to~20\% and the gun attribute to 30\%, thus dividing FB interaction as follows:~50\% cats,~30\% guns and~20\% cooking. From a total of~647 liked posts,~213 were guns and~125 cooking. In the main feed, out of~405 posts,~35 were guns and~66 cooking. There were also~7 suggested posts, out of which~4 were cooking and 2 cats. In the latest video feed, out of~65 posts,~12 were cats,~2 guns and~51 cooking. Finally, in the main video feed's~103 posts,~27 were cats and~15 cooking.

\noindent\textbf{Week 7:} We added another noise attribute that would be stronger than others. Hence, FB interaction became:~23\% cats,~23\% guns,~23\% cooking and~30\% chess. From a total of 365 liked posts,~90 were cats,~89 guns,~76 cooking and~110 chess. The main feed's 286 posts were divided as follows:~45 guns,~72 cooking and~2 chess. From 14 suggested posts,~10 were cooking and~1 chess. In the latest video feed, out of 162 posts, 18 were cats, 35 guns,~83 cooking and~22 chess. The~137 posts in the video feed were divided as follows: 25 cats,~1 guns,~9 cooking and 1 chess.

\noindent\textbf{Week 8:} The aim was to examine results, when new attributes were added without reinforcing old ones. For the first half of the week FB interaction was 100\% fishing-related and the second half 20\% fishing and 80\% bodybuilding.
\begin{itemize}
    \item \underline{First half}: Liked 235 posts about fishing. In the main feed, out of 402 posts, 207 were cats, 45 guns, 115 cooking, 4 chess and 15 fishing. Out of 7 suggested posts, 4 had to do with fishing and the others were unrelated to the user's attributes. In the latest video feed, from 190 videos, 14 were cats, 48 guns, 72 cooking, 39 chess and 18 fishing. In the main video feed, out of 148 videos, 12 were cats, 2 guns, 10 cooking, 3 chess and 1 fishing.
    \item \underline{Second half}: Liked 48 fishing posts and 181 bodybuilding posts. In the main feed, out of 423 posts, 229 were cats, 33 guns, 127 cooking, 22 fishing and 7 bodybuilding. Out of 2 suggested posts, 1 was bodybuilding and the other unrelated. In the latest video feed, out of 156 videos, 16 were cats, 9 guns, 30 cooking, 34 fishing and 72 bodybuilding. In the main video feed, out of 128 videos, 1 was cats, 2 guns, 20 cooking, 1 chess and 1 fishing.
\end{itemize}

\noindent\textbf{Week 9:} We ran \SHORT{} with 10\% cat-related traffic and the remaining with the following noise attribute layout: 20\% guns, 20\% cooking, 20\% chess, 20\% fishing, 10\% bodybuilding. From 626 liked posts, 51 were about cats, 122 guns, 130 cooking, 144 chess, 149 fishing and 29 bodybuilding. In the main feed, out of 460 posts, 199 were about cats, 51 guns, 145 cooking, 19 chess, 25 fishing and 7 bodybuilding. This time there were no suggested posts. In the latest video feed, from 154 videos, 18 had to do with cats, 14 guns, 77 cooking, 35 chess and 18 fishing. In the main video feed, from 137 videos, 25 were about cats, 1 guns, 9 cooking and 1 chess. 

\noindent\textbf{Week 10:} In the last week we ran \SHORT{} with the same parameters as in week 9: 10\% cats, 20\% guns, 20\% cooking, 20\% chess, 20\% fishing and 10\% bodybuilding. From 381 liked posts, 42 were cats, 75 guns, 96 cooking, 94 chess, 52 fishing and 22 bodybuilding. In the main feed, out of 442 posts, 160 were cats, 71 guns, 139 cooking, 30 chess, 32 fishing and 4 bodybuilding. Again, there were no suggested posts. In the latest video feed, from 133 videos, 10 were cats, 15 guns, 75 cooking, 22 chess and 12 fishing. Finally, in the main video feed, from 124 videos, 6 were cooking, 1 chess and 2 bodybuilding. 

\begin{figure}[t!]
    \centering
    \subfloat[Weekly progression of theoretical attribute strength]{\label{fig:theoreticalA}{\includegraphics[width=0.40\textwidth]{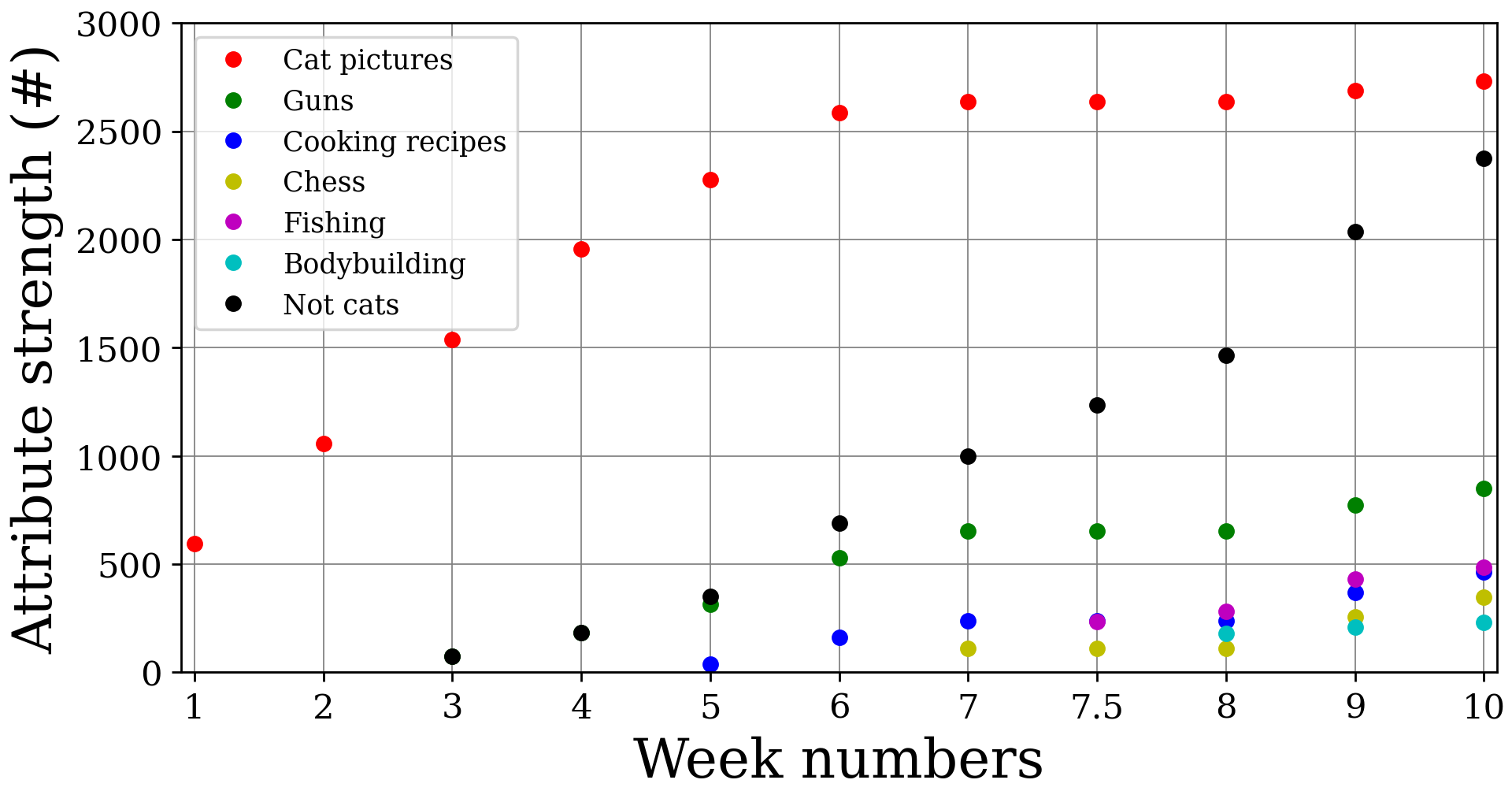}}}%
    \qquad
    \subfloat[Ratio of weekly liked posts]{\label{fig:theoreticalB}{\includegraphics[width=0.40\textwidth]{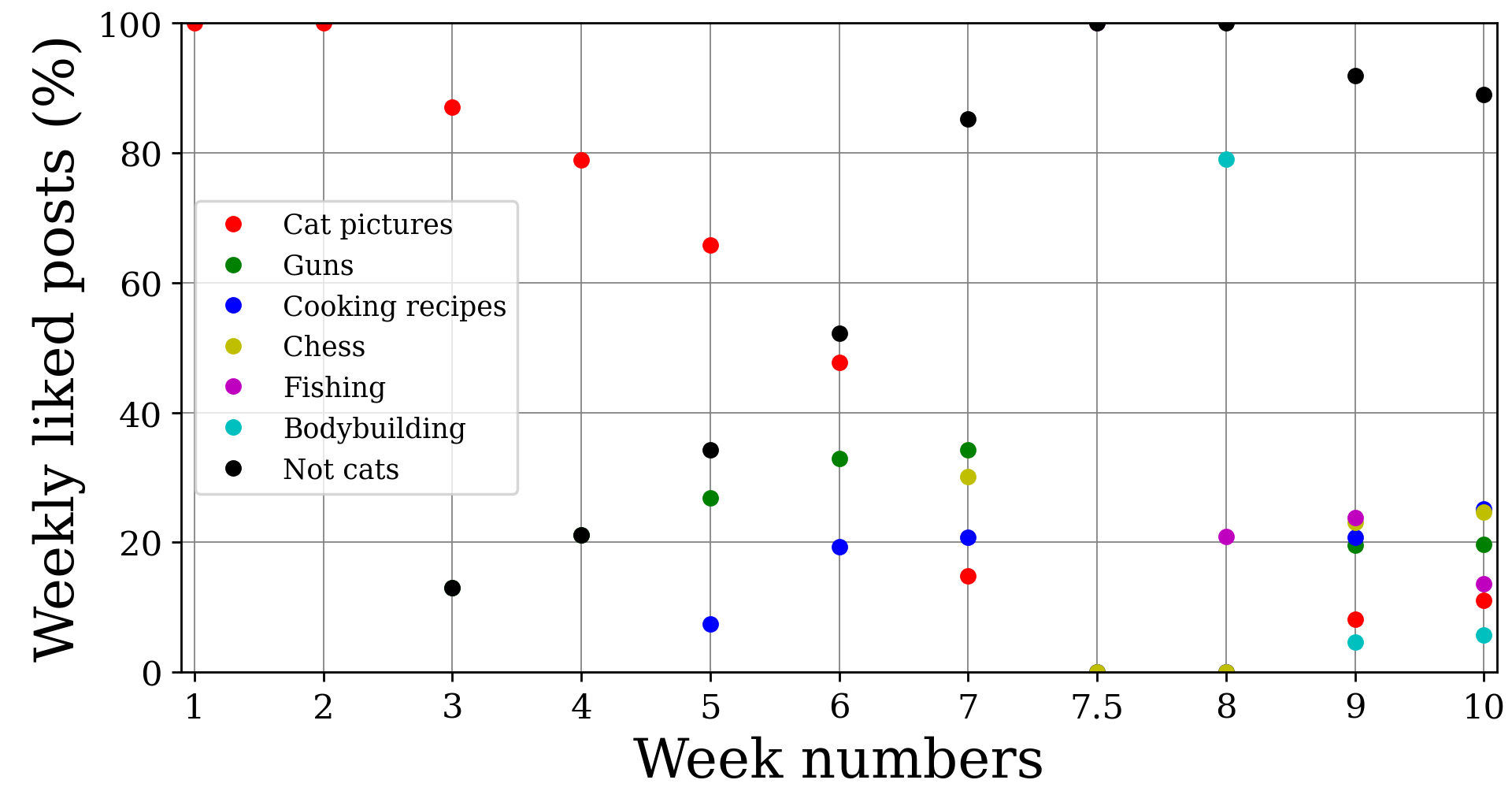} }}%
    \caption{The total amount of posts liked and the ratio of posts liked per week.}
    \label{fig:theoretical}
\end{figure}

The total amount of posts liked on a weekly basis for each attribute (attribute strength), is shown in \autoref{fig:theoreticalA}. The week number is noted on the horizontal axis and the attribute strength (total amount of posts liked) on the vertical axis. As the figure indicates, even on week 10, the `cat' attribute strength outweighs all others combined, since the attribute remained reinforced even when said reinforcement decreased over time. 
\autoref{fig:theoreticalB} represents the ratio of posts. Here, the ratio is calculated using the posts liked on a specific week, omitting those of previous weeks. This time, the attribute strength on the vertical axis stands for the percentage of liked posts for each attribute.

\begin{figure}[ht]
    \centering
    \subfloat[The percentage of posts for each attribute from liked pages in the main feed.]{\label{fig:effective1A}{\includegraphics[width=0.35\textwidth]{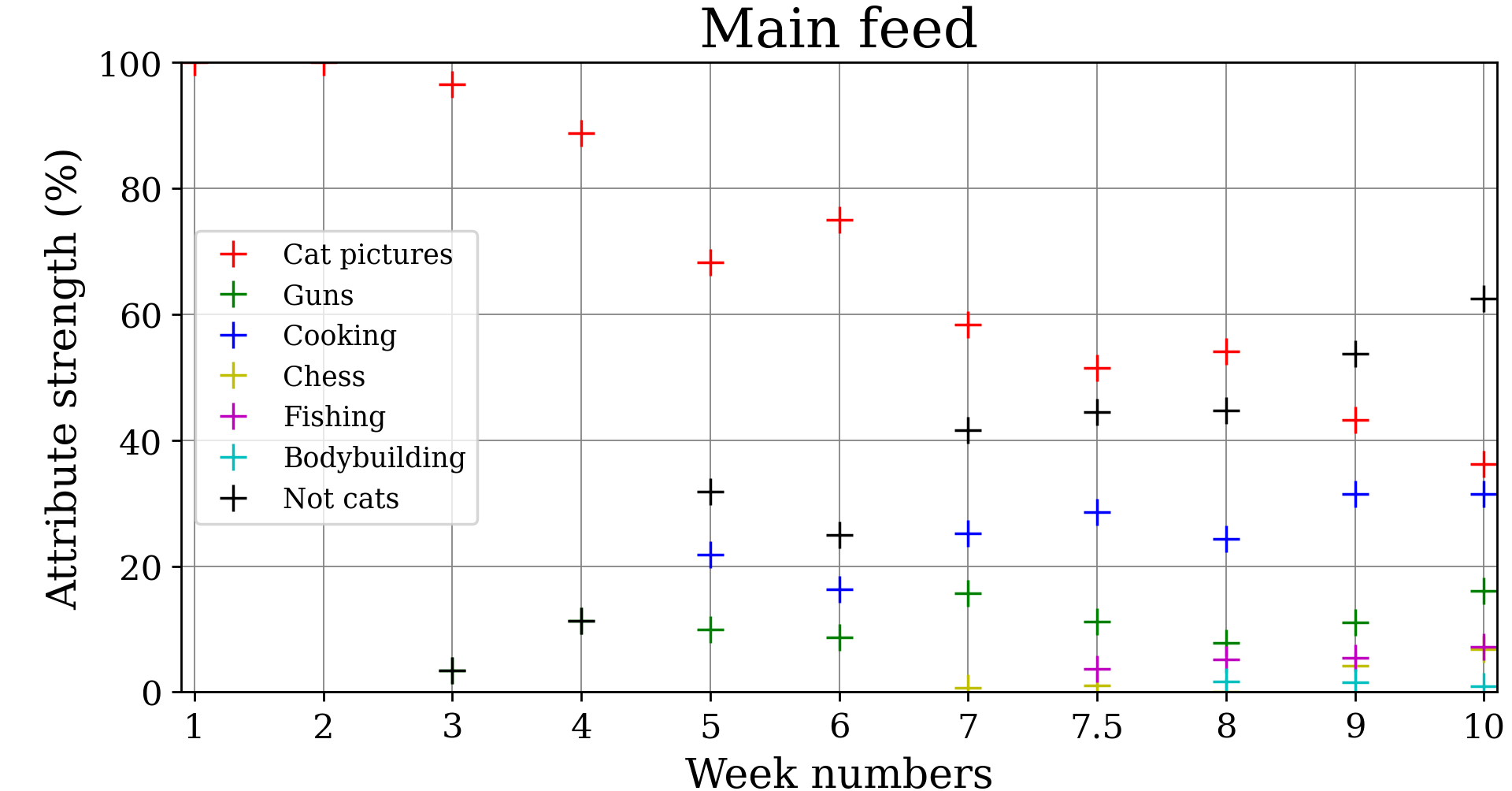}}}%
    \qquad
    \subfloat[The percentage of recommended, suggested and sponsored posts for each attribute in the main feed from pages not liked by the user nor the \BOT{} .]{\label{fig:effective1B}{\includegraphics[width=0.35\textwidth]{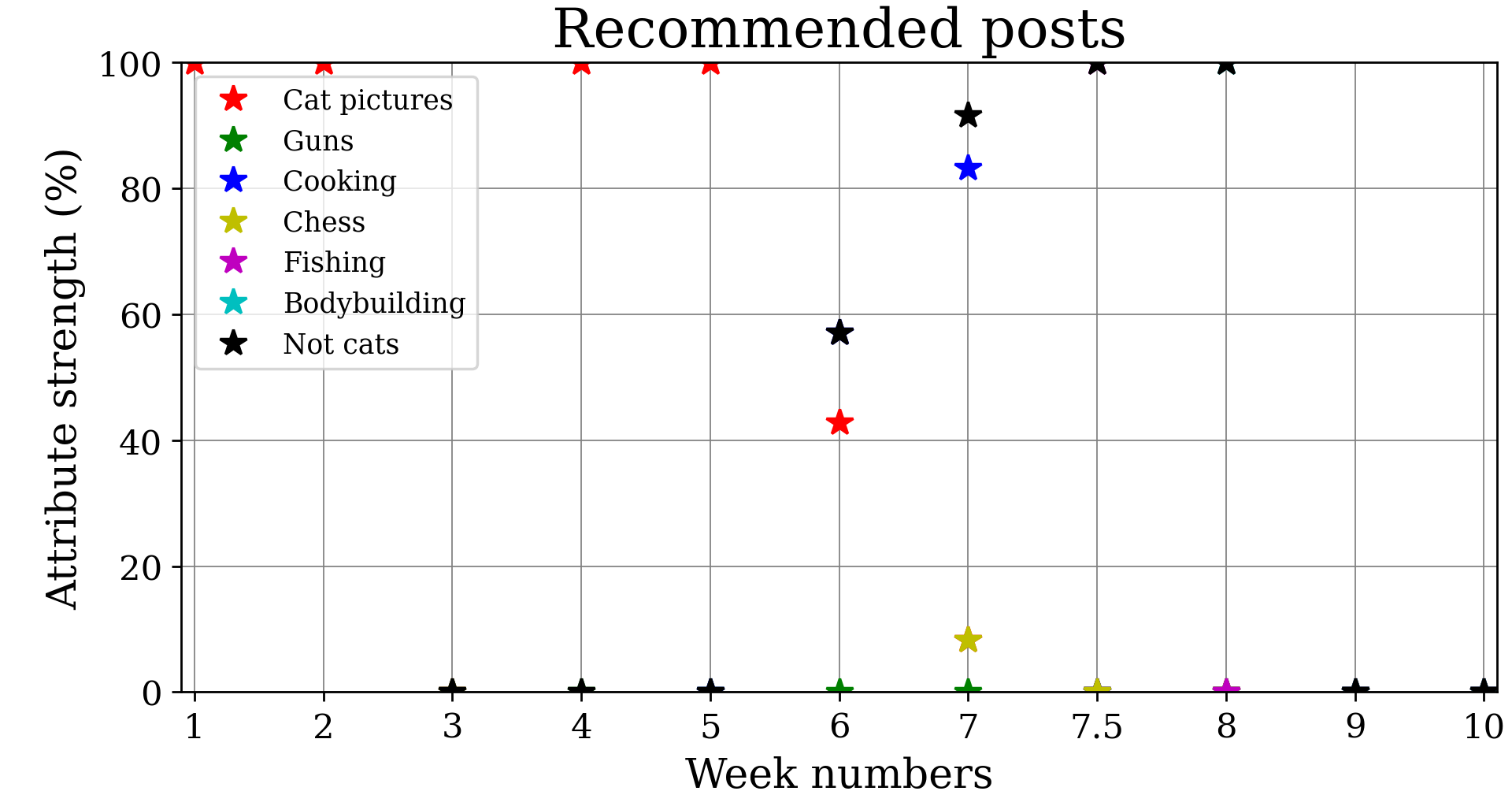} }}%
    \\
    \subfloat[The percentage of video posts for each attribute from the latest video feed.]{\label{fig:effective1C}{\includegraphics[width=0.35\textwidth]{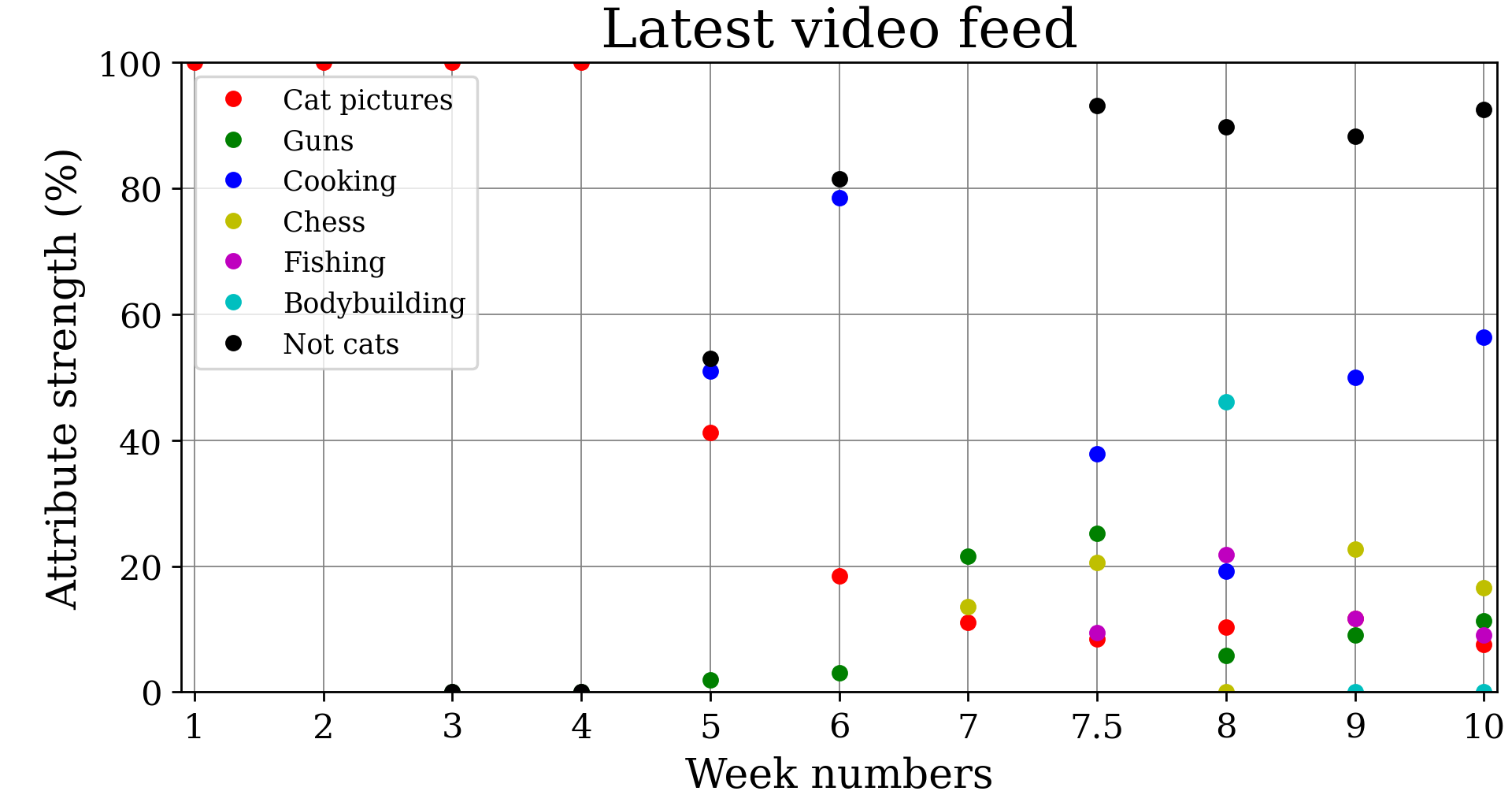} }}%
    \qquad
    \subfloat[The percentage of each attribute of video posts from the main video feed.]{\label{fig:effective1D}{\includegraphics[width=0.35\textwidth]{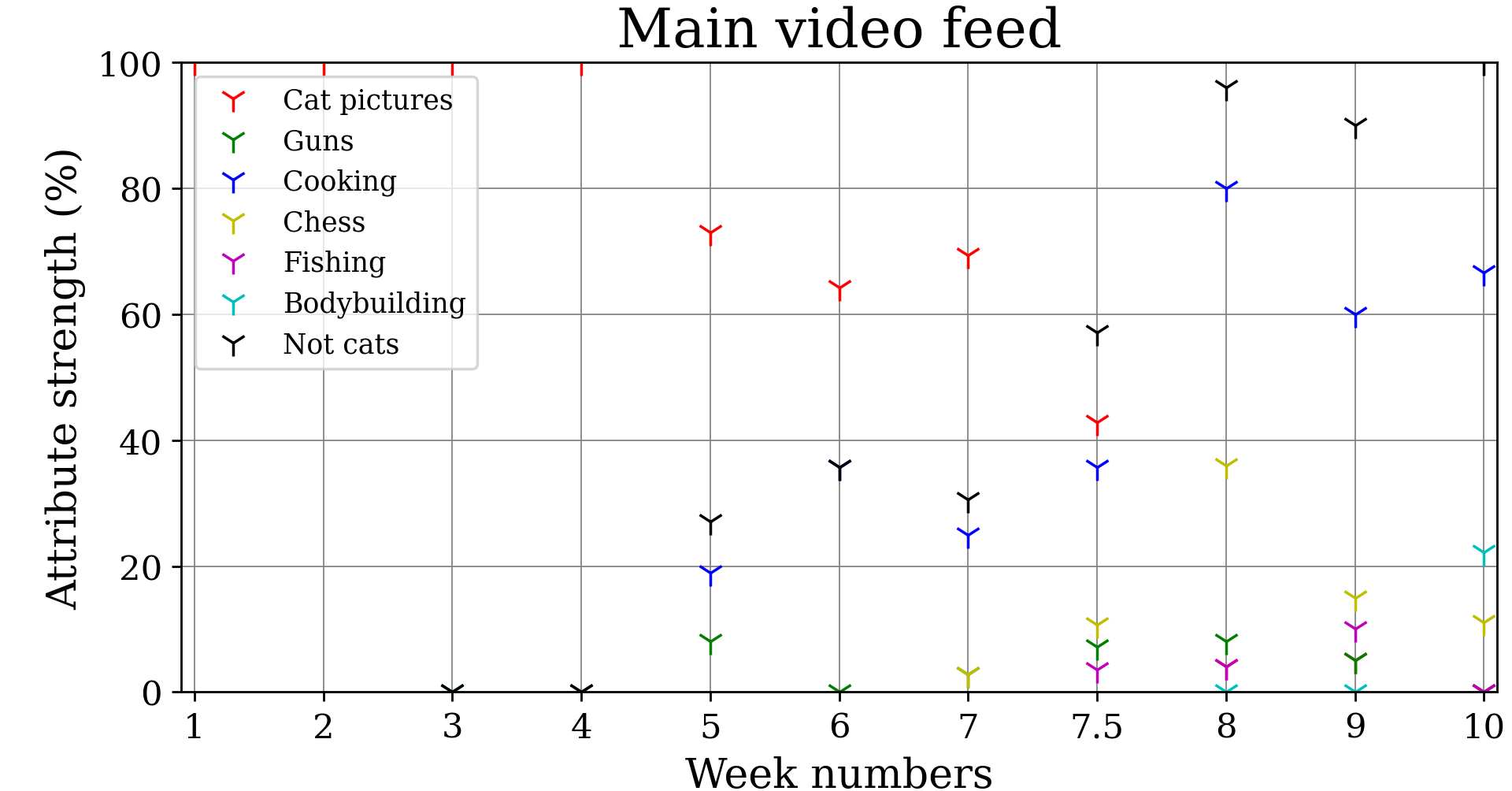}}}%
    \caption{Effective attribute strength variables with combined noise}%
    \label{fig:effective1}
\end{figure}

Next, we present the results of each variable for the effective attribute strength. The main feed, recommended posts, latest video feed and main video feed are represented in \autoref{fig:effective1}. 

We can, now, compare results between \autoref{fig:effective1} and \autoref{fig:theoreticalB}: on weeks~5 to~8, noise-effective attribute strength variables approached real variables. \autoref{fig:theoreticalB} shows that around week~6, there are more noise-related likes than real likes. Consequently, FB's recommendations 
show more noise-related content as we can see from \autoref{fig:effective1}. In the first~4 weeks, \autoref{fig:effective1C} and \autoref{fig:effective1D} show no relation to noise attributes. We thus conclude that 20\% noise is not enough to change said variables. Also, \autoref{fig:effective1B} shows that in a few weeks' time, there were no recommended/suggested posts in the main feed (weeks 3, 9 and 10).

To avoid confusion in \autoref{fig:effective1} we must clarify that in the main video feed \autoref{fig:effective1D} and the recommended, suggested and sponsored posts \autoref{fig:effective1B}, the FB content is derived from pages not liked by the user. The content is both user attribute-related and unrelated. It is assumed that the unrelated content is presented by FB because of other features in their recommendation systems e.g.\ users who liked X also liked Y. Their recommendation algorithms are not open source, hence their mode of operation is concealed. Due to this, our results are based on content exclusively related to user attributes.


\subsection{Privacy Results}
\label{subsec:priv_results}
\begin{figure}[ht]
    \centering
    \subfloat[ ][Theoretical privacy]{\label{fig:privacyA}{\includegraphics[width=0.35\textwidth]{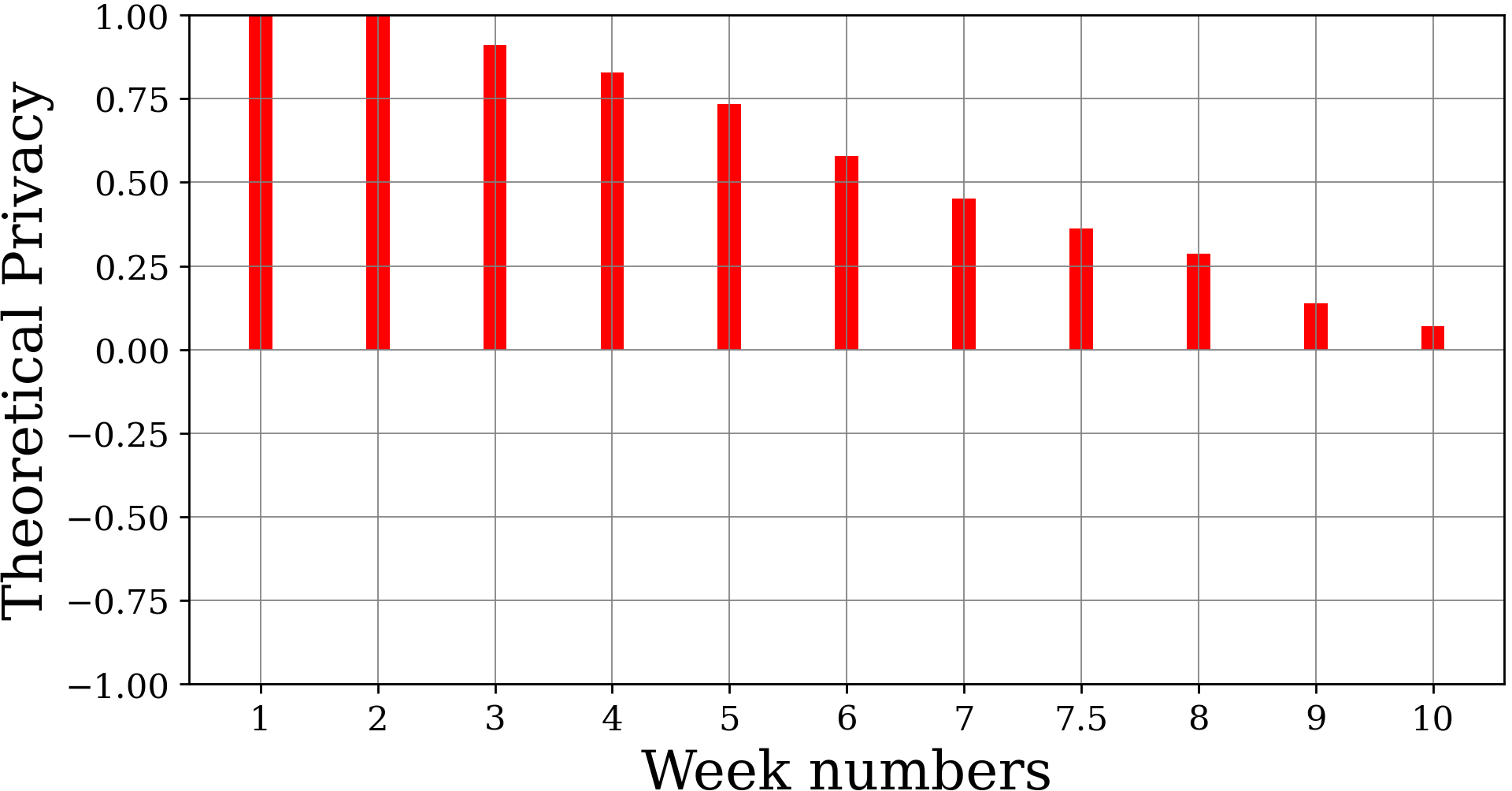}}}%
    \qquad
    \subfloat[ ][Effective privacy]{\label{fig:privacyB}{\includegraphics[width=0.35\textwidth]{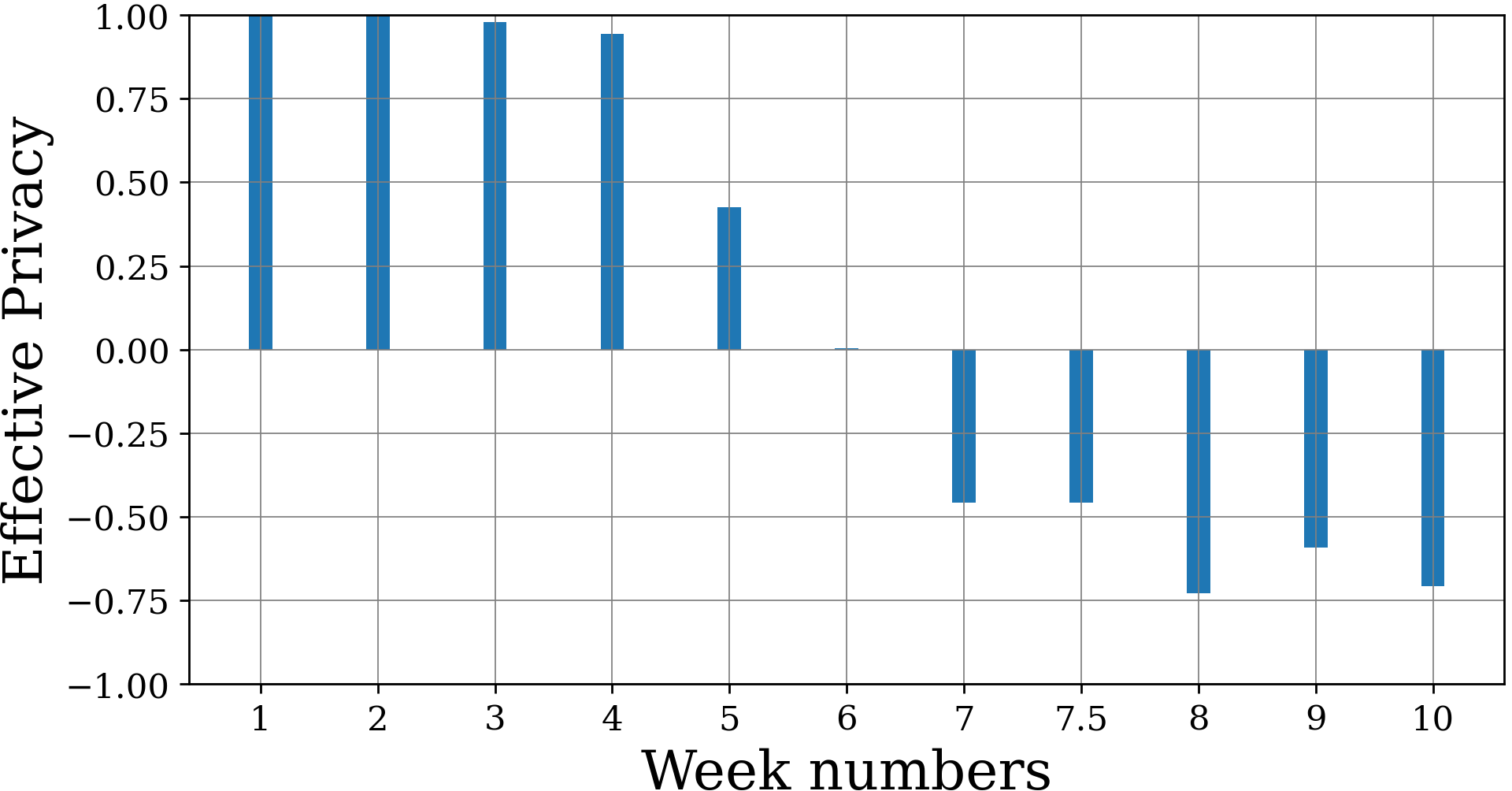} }}%
    \caption{Theoretical and Effective privacy results}
    \label{fig:privacy}
\end{figure}

Based on the definitions described in \autoref{sec:QuantifyPrivacy}, we calculated each week's Theoretical (\autoref{fig:privacyA}) and Effective Privacy (\autoref{fig:privacyB}) values. During the first two weeks, we built the user's real attributes and added increasing noise to render FB's noise feed equal to the real. 



Notably, the Effective Privacy in week 6 (50\% noise) is close to 0. 
Once the amount of real traffic generated by users equals the amount of noise traffic, users achieve privacy. The theoretical real attribute strength outweighs the combined noise attribute strengths even after 10 weeks, as shown in \autoref{fig:theoreticalA}. This explains the difference between the Theoretical and Effective Privacy values and shows that FB emphasises on the user's recent interests, suggesting a ``time of like" variable in its recommendation systems. This also proves that the Effective Privacy is a more accurate way of measuring privacy on a SN.

We added more noise in week~7 and saw a small decrease in the Effective Privacy value -- i.e.\ the account became more private. During week 8, we stopped reinforcing the real attribute to simulate what would happen if the user took a break from FB, while the \BOT{} ran. We noted significant decrease in the Effective Privacy value. Finally, in weeks 9 and 10, we simulated a rarely active user combined with \BOT{} background activity (90\% noise). The Effective Privacy value increased as the real attribute was re-enforced again in week 9, while the Effective Privacy value decreased again during week 10.


\subsection{Real Account Results}
When evaluating \SHORT{} extended features on real existing accounts, a different approach was used as compared to that used for the dummy account. The theoretical privacy takes into account all the likes that a user did during their entire history with FB, which is unfeasible to obtain and categorise into keywords. To this end, when analyzing the results, we considered the feed from pages related to the noise keywords as our noise attribute. For the real data, we simply considered the feeds 
unrelated to our chosen noise keyword. To this end, we ran our experiments for 4 weeks on two existing accounts (account A and account B). Account A was set to like an average of 27 posts per day, while account B was set to like 54 posts per day.

\smallskip

\begin{figure}[ht]
    \centering
    \subfloat[Keywords and their respective amount of liked posts, liked pages and watched videos]{\label{fig:eugeneA}{\includegraphics[width=0.35\textwidth]{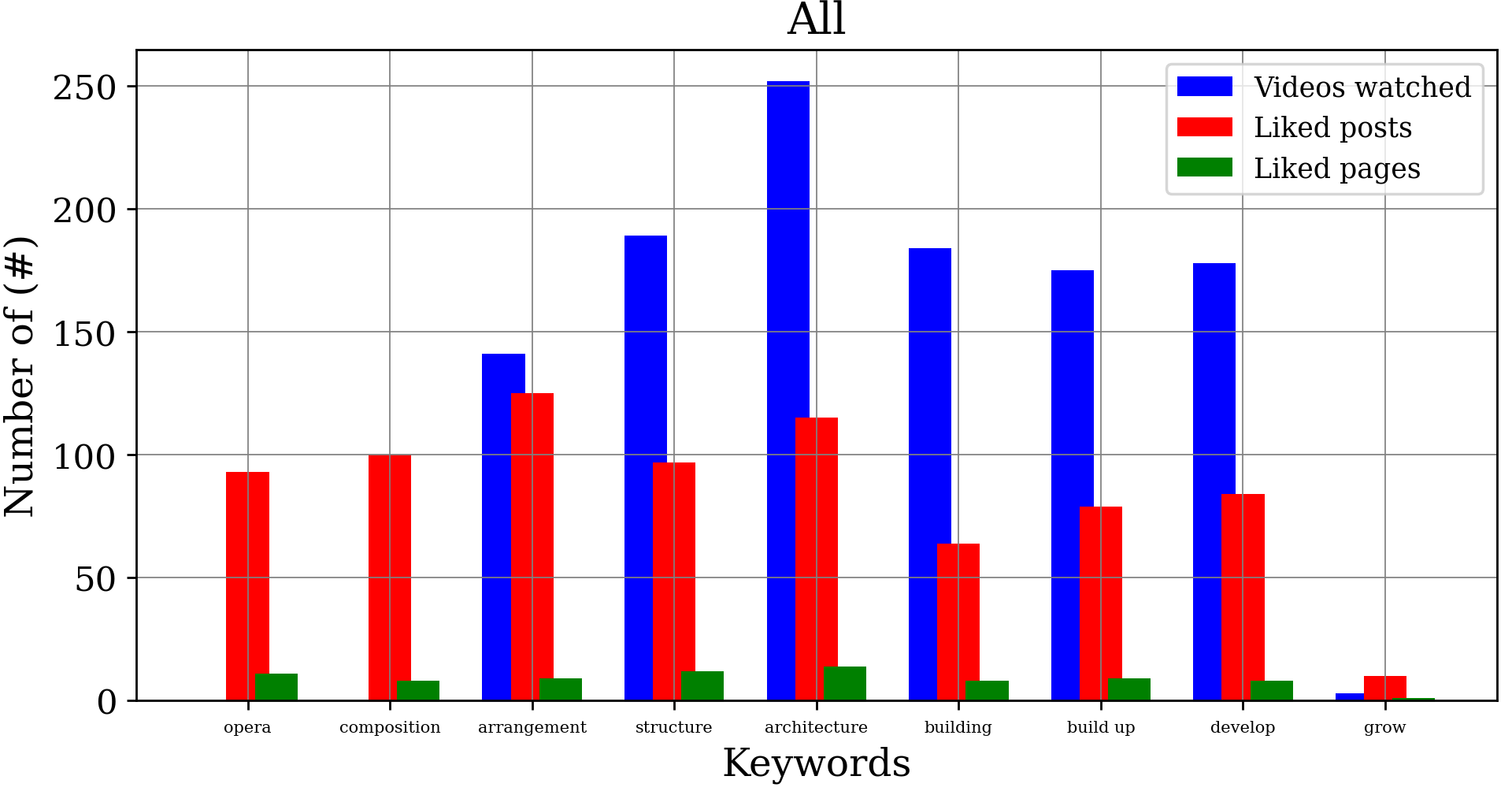}}}%
    \qquad
    \subfloat[Keywords and their respective amount of liked pages]{\label{fig:eugeneB}{\includegraphics[width=0.35\textwidth]{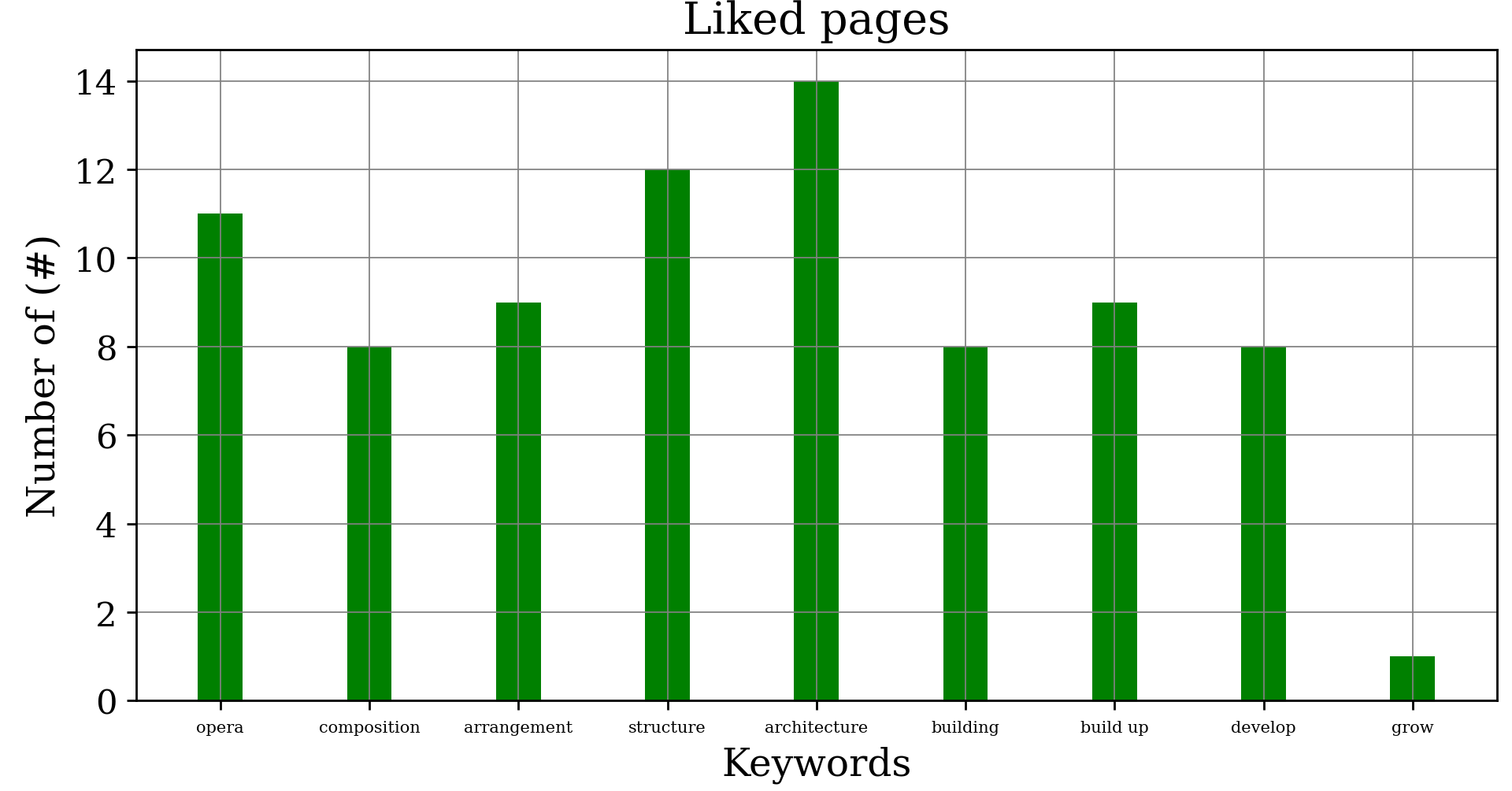} }}%
    \\
    \subfloat[Keywords and their respective amount of liked posts]{\label{fig:eugeneC}{\includegraphics[width=0.35\textwidth]{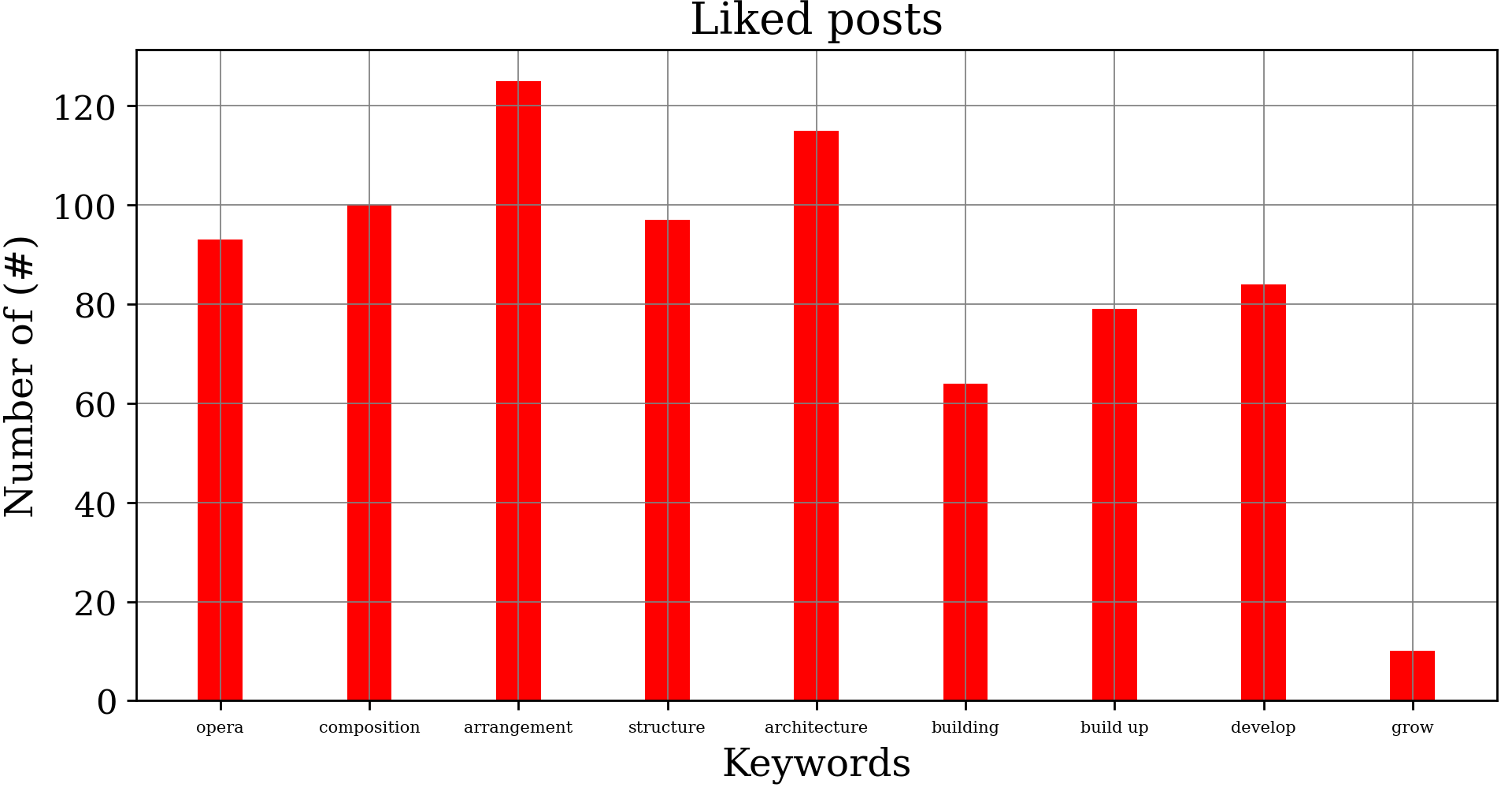} }}%
    \qquad
    \subfloat[Keywords and their respective amount of watched videos]{\label{fig:eugeneD}{\includegraphics[width=0.35\textwidth]{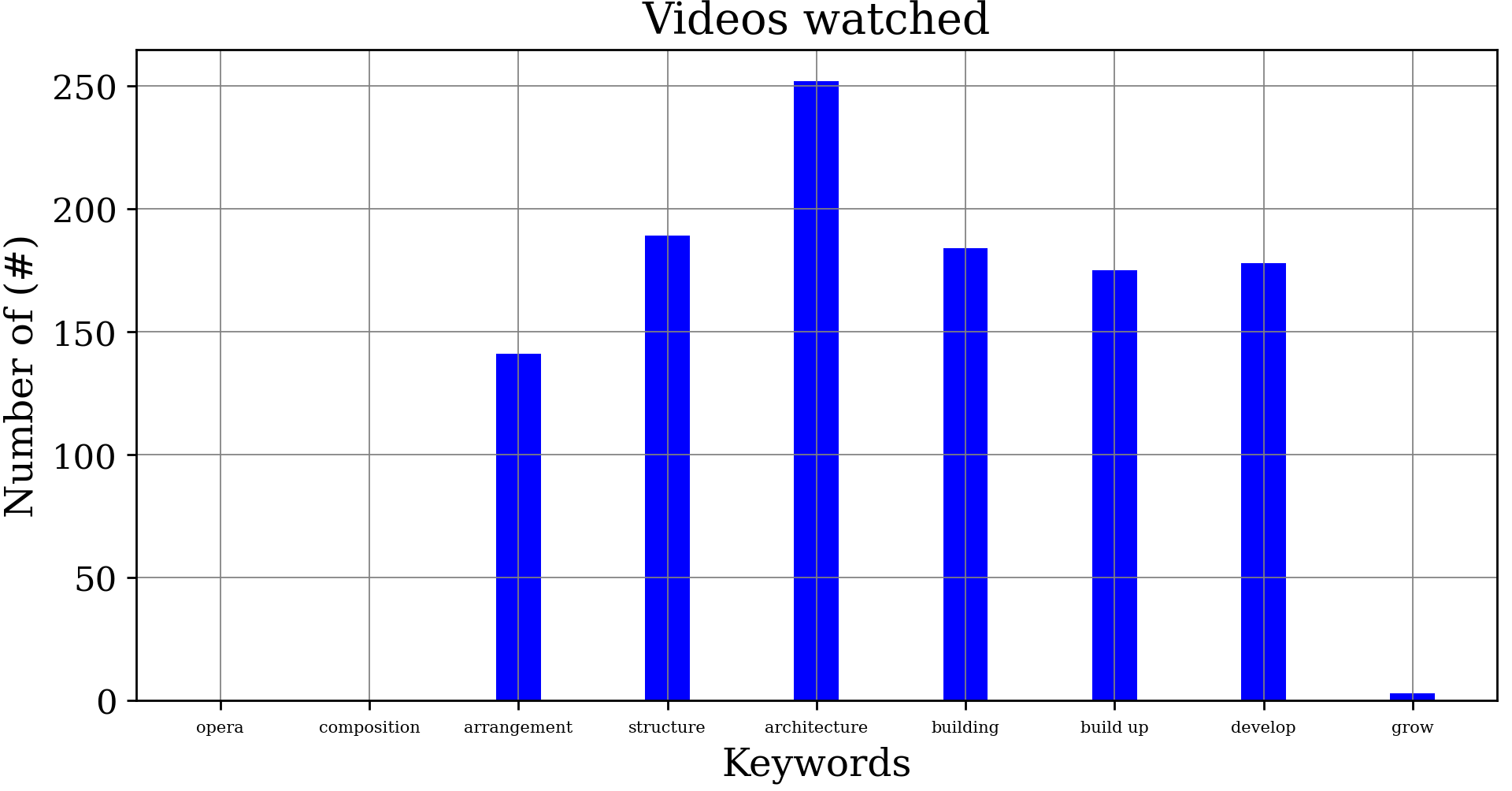}}}%
    \caption{Keyword statistics  of account A.}%
    \label{fig:eugene}
\end{figure}

\noindent\textit{\textbf{Account A:}} During the 4 week period, the \BOT{} liked~754 posts from~79 pages and watched~1122 videos. The chosen seed keyword was `opera' after which the \BOT{} generated other related noise keywords. These keywords, along with their respective amount of posts, pages and videos can be seen in \autoref{fig:eugene}. During the first week of our evaluations, the \BOT{} used the first two keywords: `opera' and `composition'. We then analyzed the results, and again observed a complete absence of noise from the FB feeds (this eventually led to the development of the video watching and link clicking features). The extended version of \SHORT{} run for the subsequent~3 weeks. 

\begin{figure}[ht]
    \centering
    \subfloat[Raw representation]{\label{fig:eugeneresA}{\includegraphics[width=0.35\textwidth]{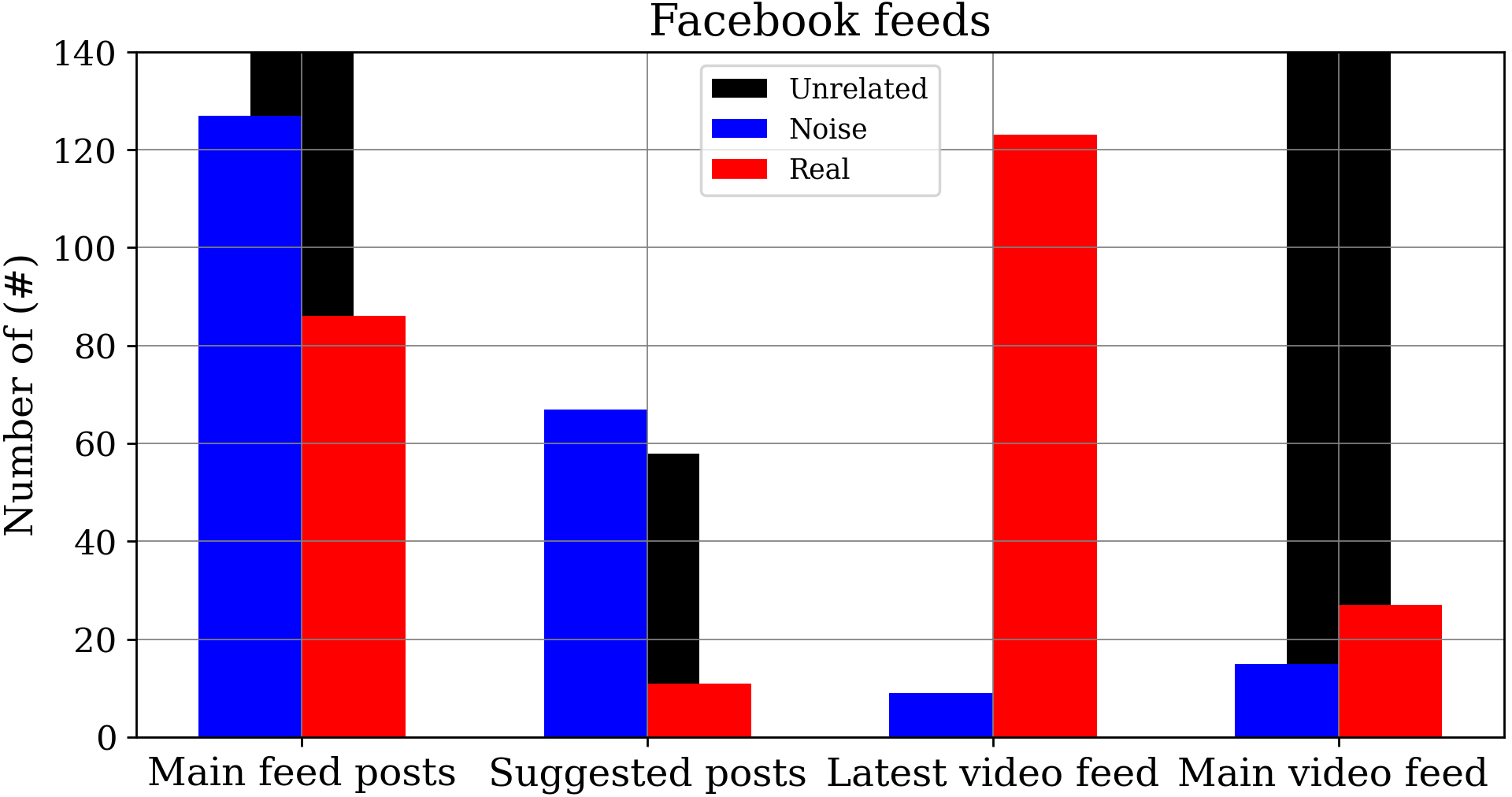}}}%
    \qquad
    \subfloat[Percentage representation]{\label{fig:eugeneresB}{\includegraphics[width=0.35\textwidth]{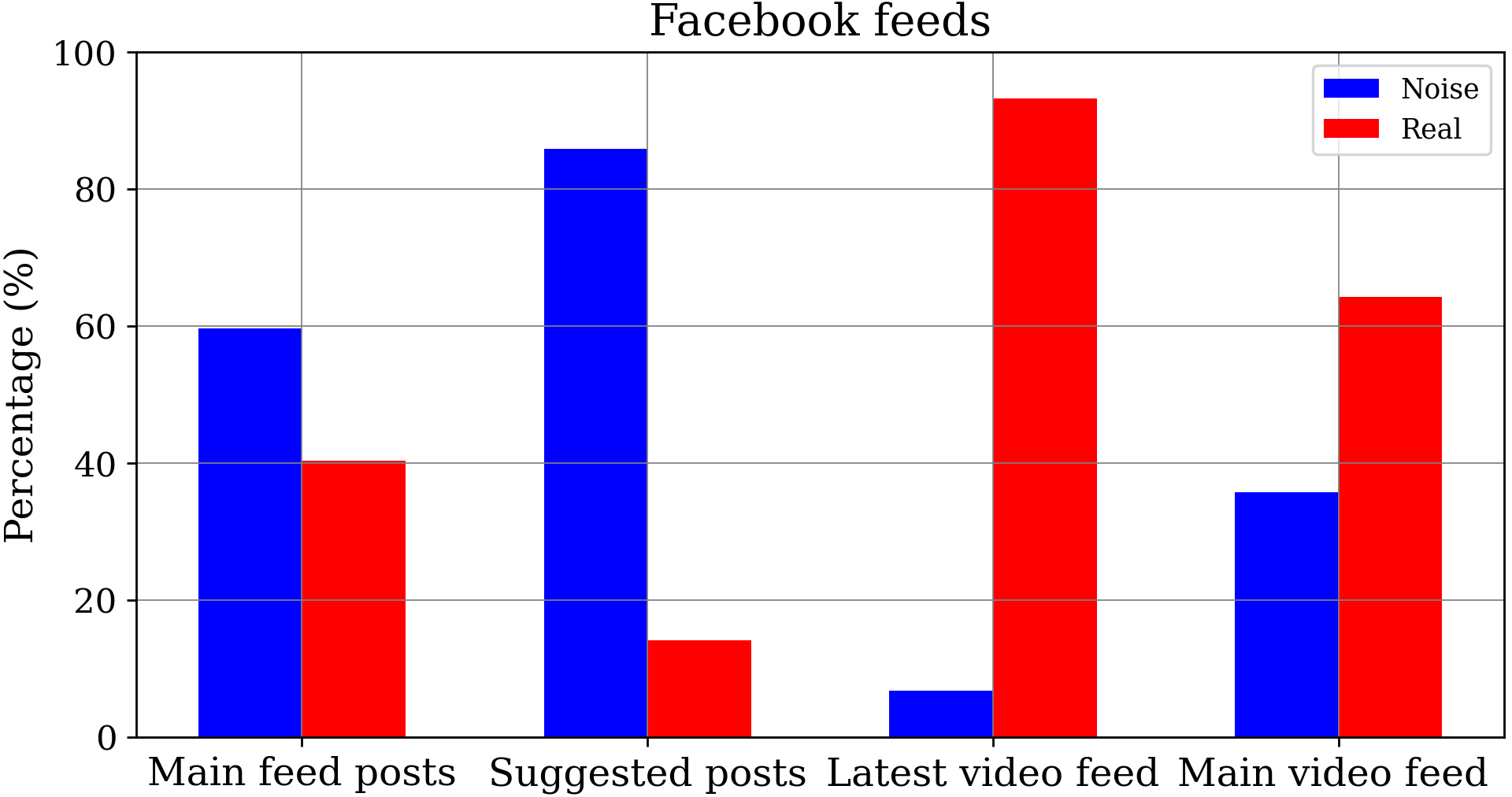} }}%
    \caption{Graphical representation of the collected data from account A.}
    \label{fig:eugeneres}
\end{figure}

We then analyzed the different FB feeds. In the main feed, out of~596 posts,~86 were related to real interests and~127 were noise related. From~136 suggested posts, 11 were based on real interests,~67 were based on noise and~58 seemed to be related only to location (local grocery advertisements, etc.). In the latest video feed, from~132 videos,~123 were related to real interests and~9 to noise. Finally, in the main video feed, out of~300 videos,~27 were real interest related,~15 were noise related and the rest seemed unrelated to noise or real interests. 
These results are represented graphically in \autoref{fig:eugeneres}. \autoref{fig:eugeneresA} shows the exact number of real and noise data (the unrelated bar is out of bounds as it it not used in the calculations) and \autoref{fig:eugeneresB} shows the percentage of real and noise data.

\begin{figure}[ht]
    \centering
    \subfloat[Keywords and their respective amount of liked posts, liked pages and watched videos]{\label{fig:alexA}{\includegraphics[width=0.35\textwidth]{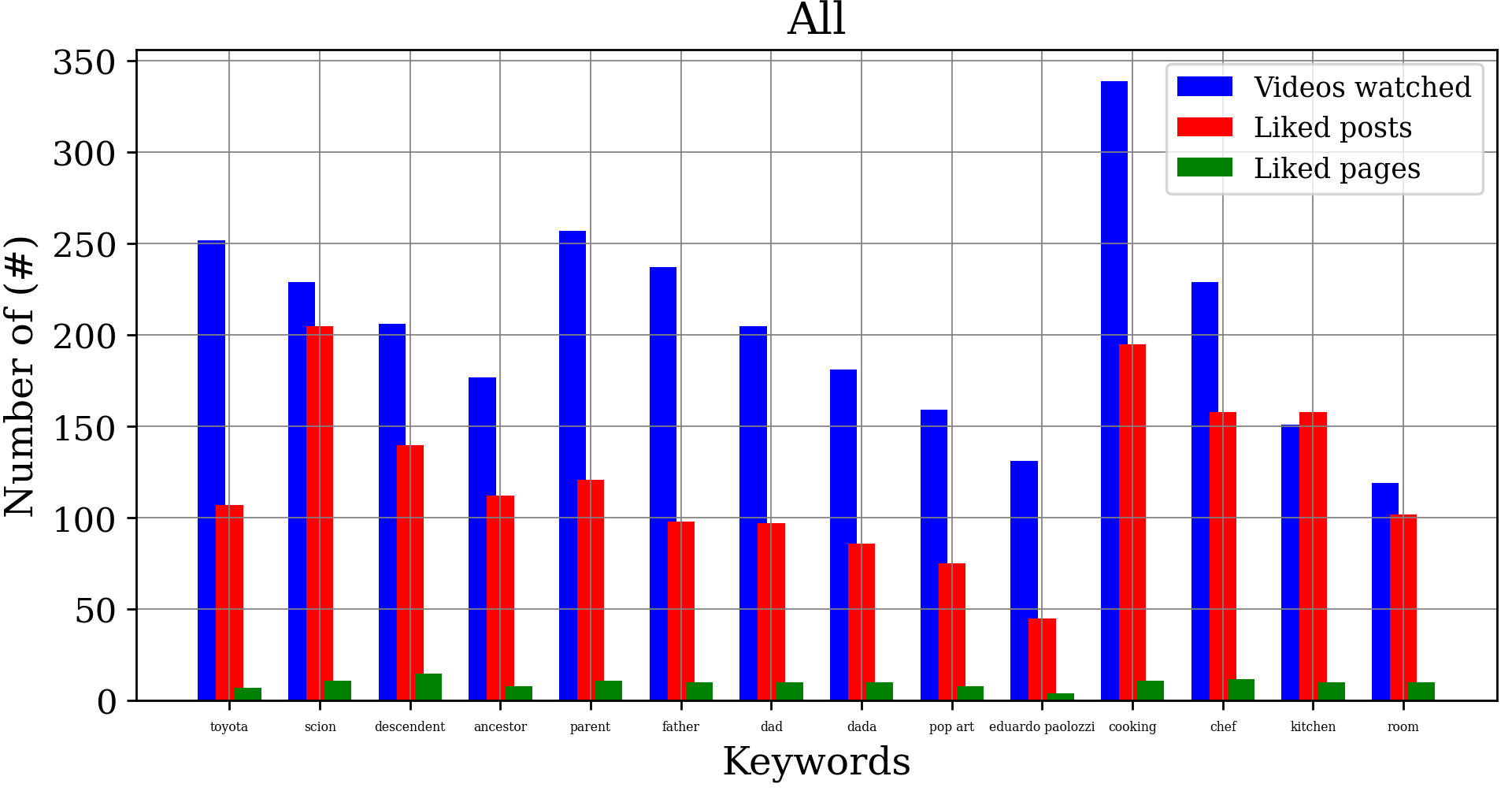}}}%
    \qquad
    \subfloat[Keywords and their respective amount of liked pages]{\label{fig:alexB}{\includegraphics[width=0.35\textwidth]{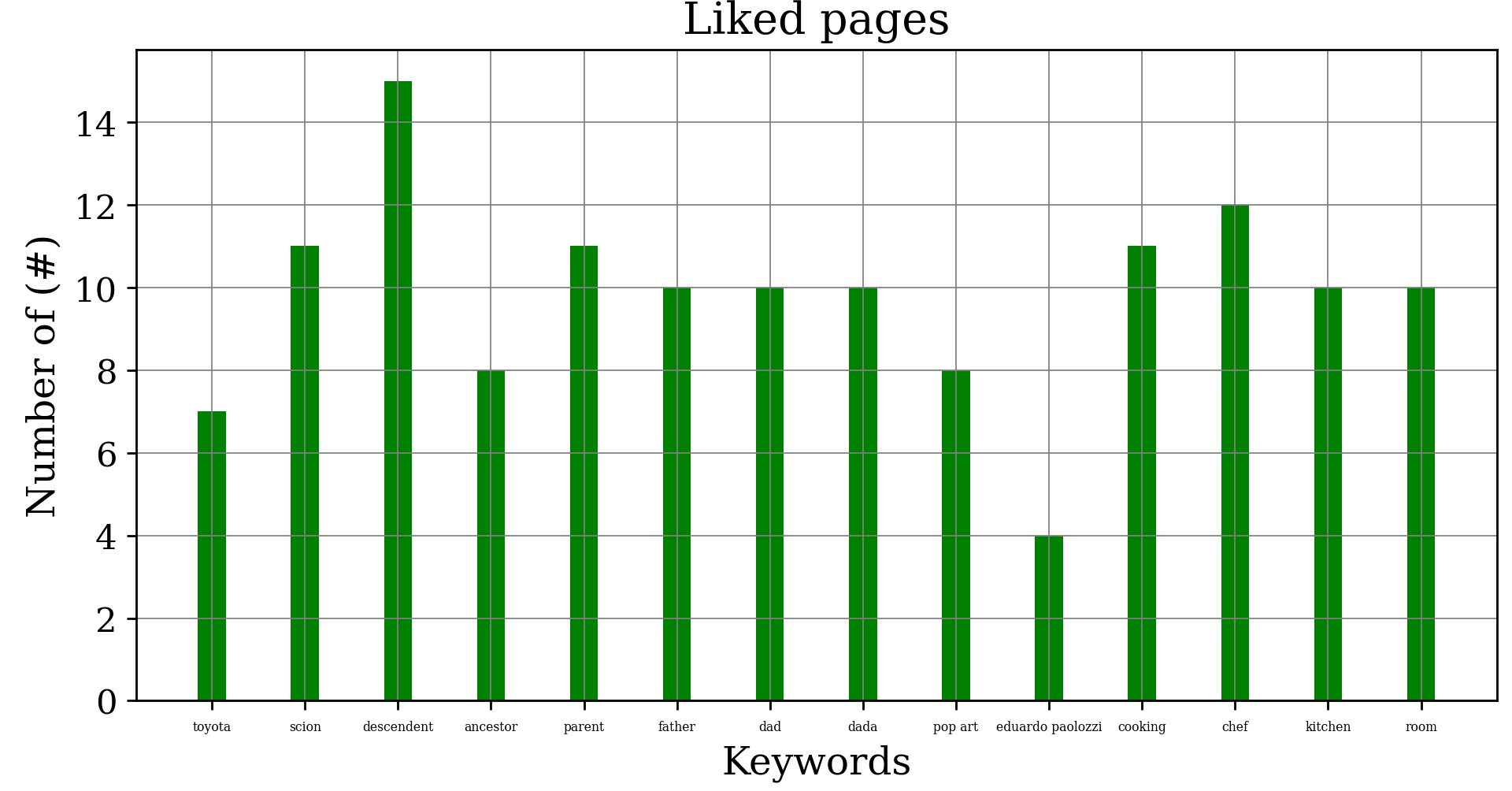} }}%
    \\
    \subfloat[Keywords and their respective amount of liked posts]{\label{fig:alexC}{\includegraphics[width=0.35\textwidth]{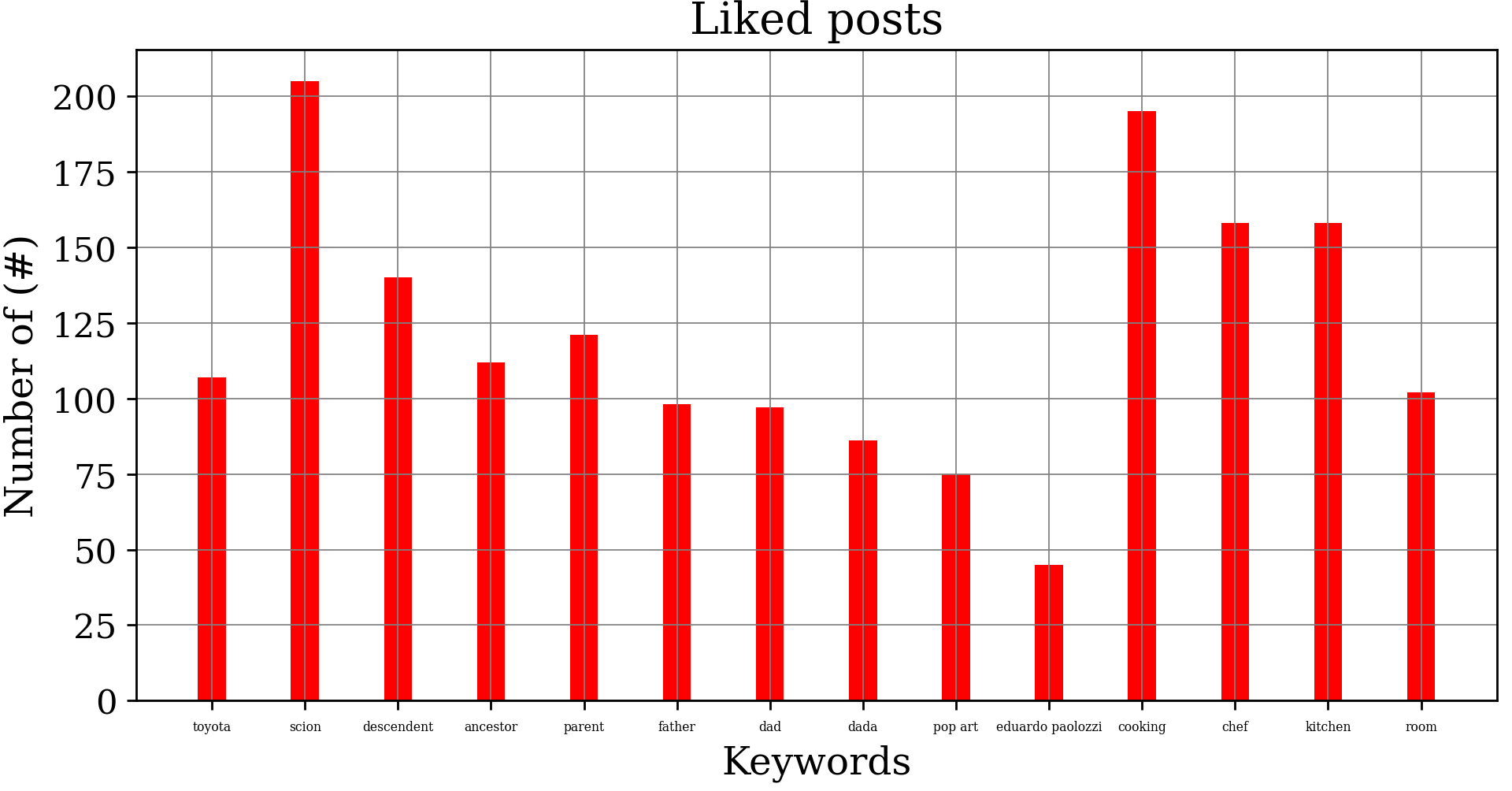} }}%
    \qquad
    \subfloat[Keywords and their respective amount of watched videos]{\label{fig:alexD}{\includegraphics[width=0.35\textwidth]{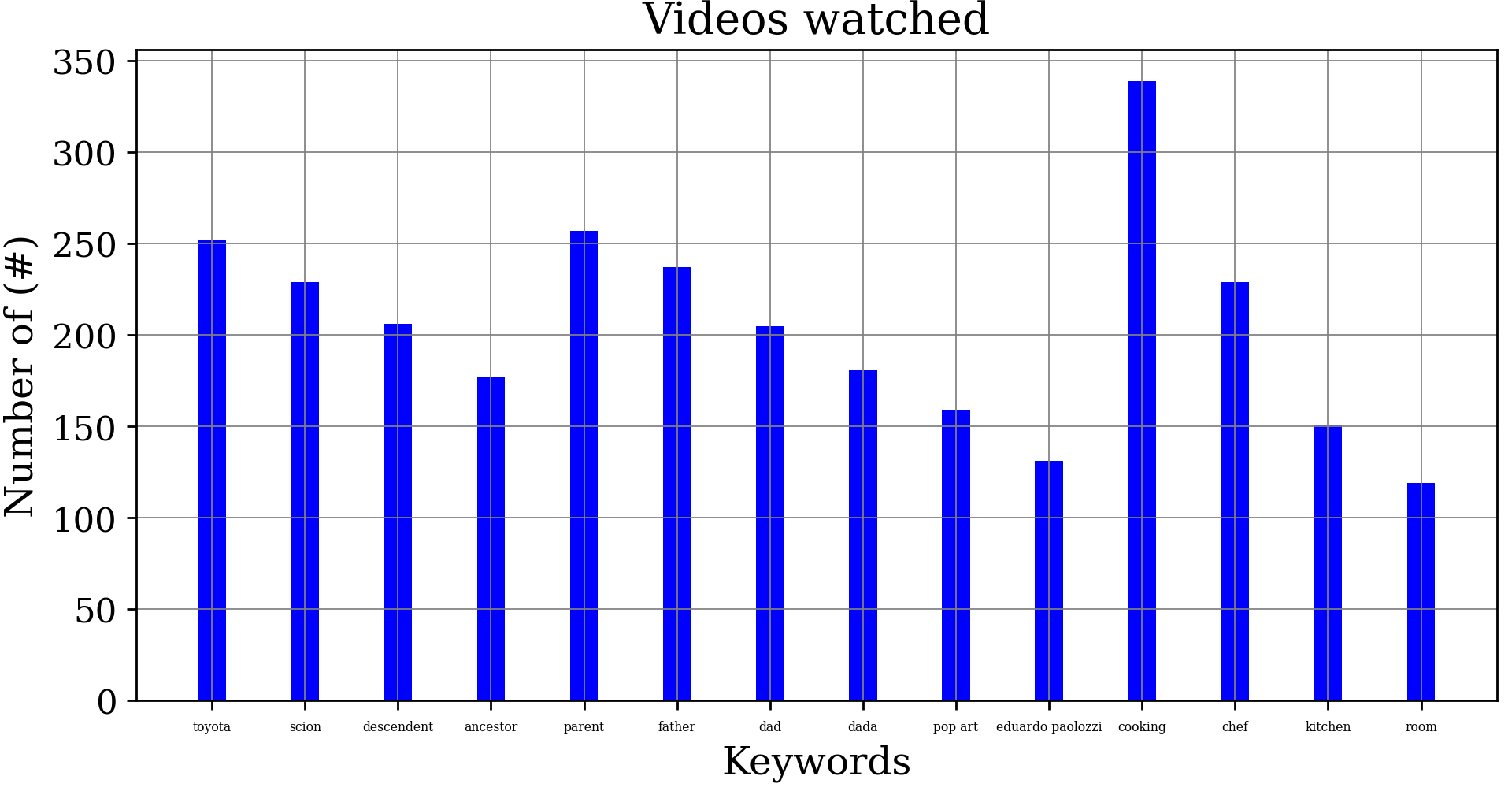}}}%
    \caption{Keyword statistics of account B.}%
    \label{fig:alex}
\end{figure}

\begin{figure}[ht]
    \centering
    \subfloat[Raw representation]{\label{fig:alexresA}{\includegraphics[width=0.35\textwidth]{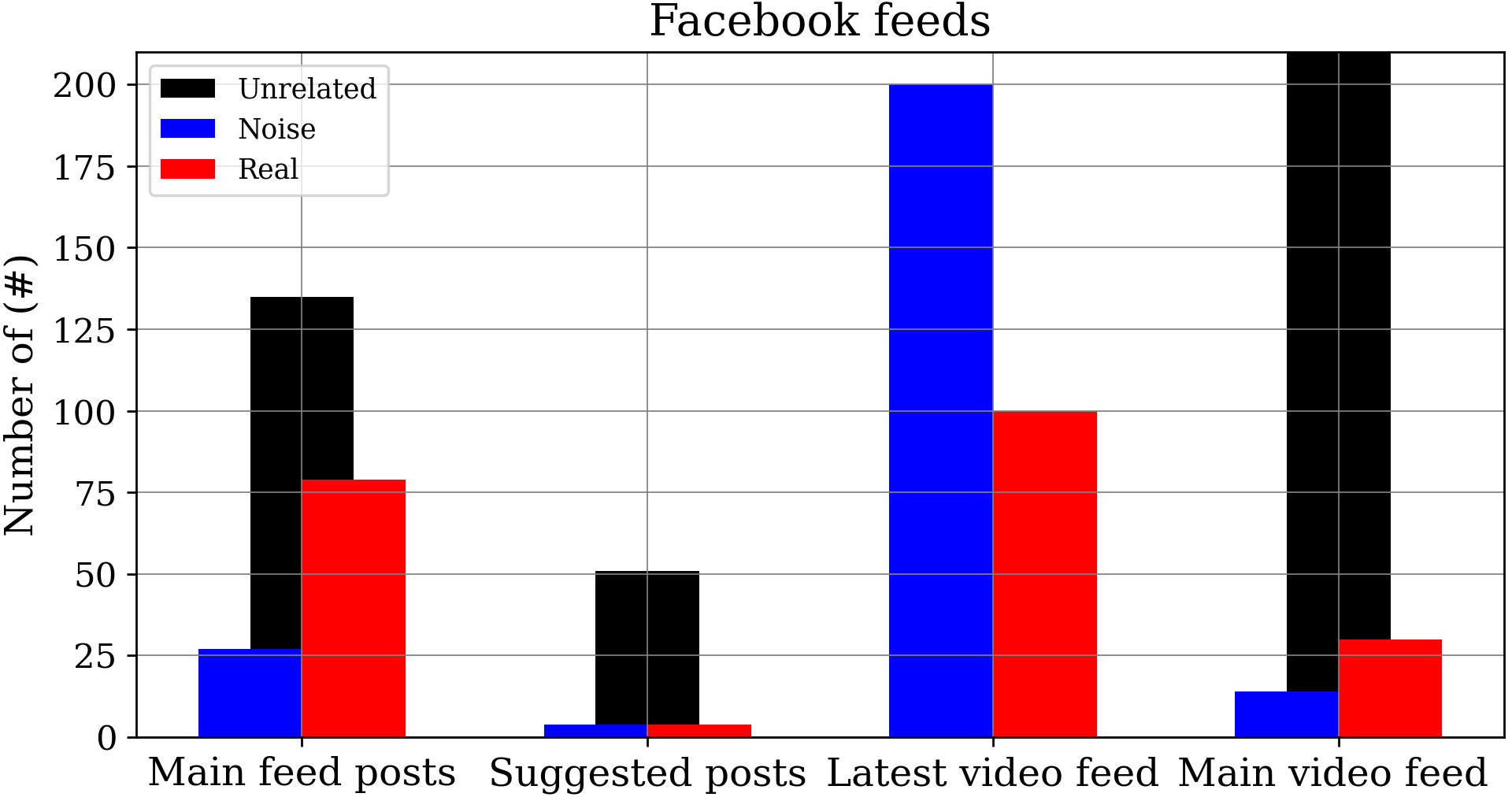}}}%
    \qquad
    \subfloat[Percentage representation]{\label{fig:alexresB}{\includegraphics[width=0.35\textwidth]{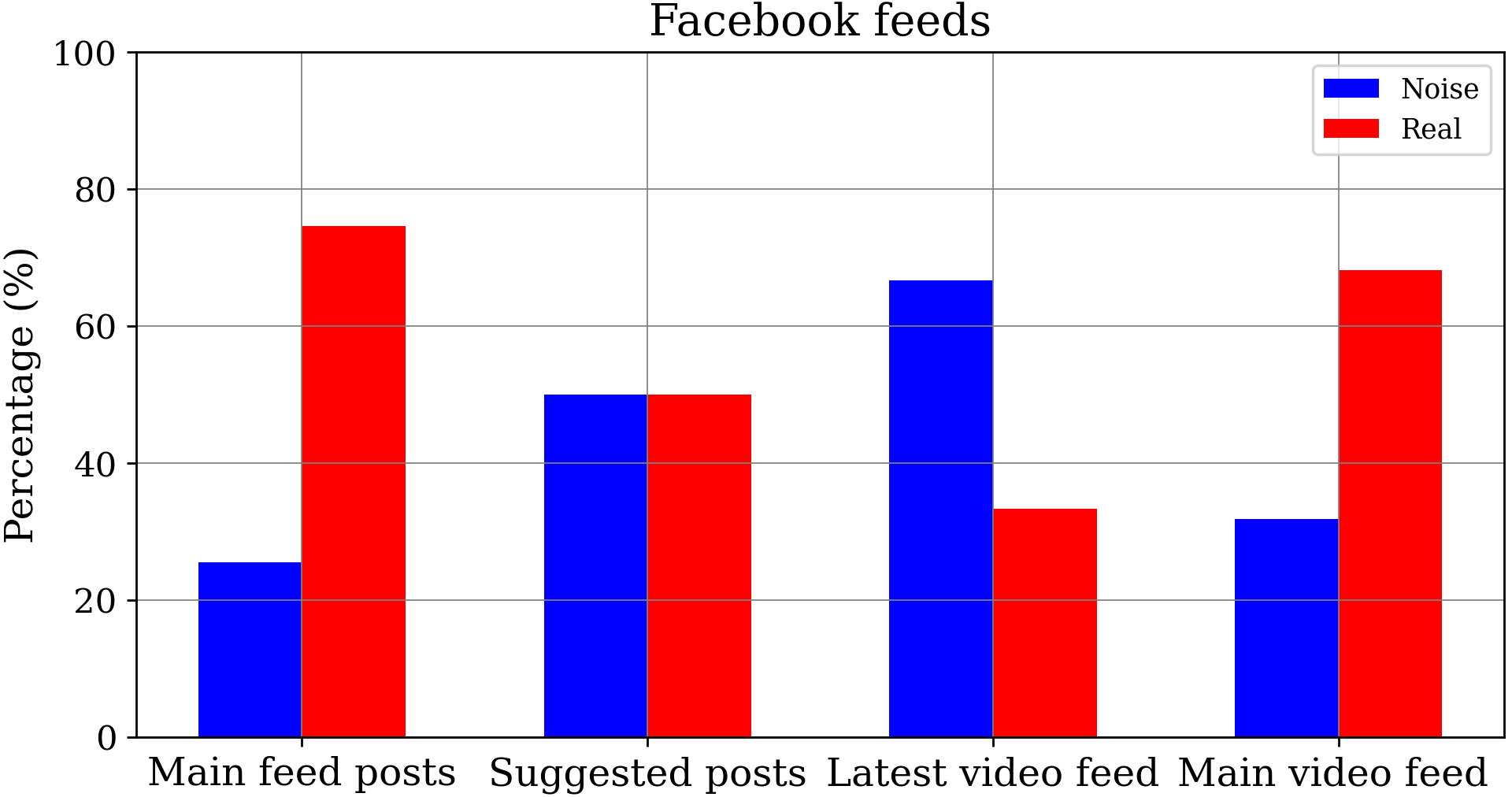} }}%
    \caption{Graphical representation of the collected data from account B.}
    \label{fig:alexres}
\end{figure}

\noindent\textit{\textbf{Account B:}} For this account, the \BOT{} liked 1518 posts from 129 pages and watched 2871 videos. The chosen seed keyword was `toyota' and the keyword statistics are shown in \autoref{fig:alex}. With this account, the extended \SHORT{} was used from the beginning. Once completed, we analyzed the different FB feeds. In the main feed, out of 300 posts, 79 were real interest related and 27 were noise related. From 59 suggested posts, 4 were based on real interests, 4 were based on noise and 51 seemed to be related only to location (local grocery advertisements, data carriers, etc.). In the latest video feed, from 300 videos, 100 were related to real interests and 200 were related to noise. Finally, in the main video feed, out of 300 videos 30 were real interest related, 14 were noise related and the rest seemed unrelated to noise or real interests. The results are also represented graphically in \autoref{fig:alexres}.

\noindent 

It should be noted that analyzing the feed of a real account can be a tedious process as every post has to be manually inspected. Implementing a script to automate this task would be very challenging as there is no way it could distinguish between posts from pages and posts from friends and groups. Additionally, for the FB feed that came as suggested (suggested posts and main video feed), it was harder to understand what was related to the real or noise data. With the old accounts there are significantly more variables to consider such as friends, devices, locations etc.\ as opposed to a new account 
that exists in a controlled environment. Hence, the  results should be treated as an approximation. 

\noindent\textbf{Observations:} We compared the percentage representation of the FB feeds from the 2 accounts: \autoref{fig:eugeneresB} and \autoref{fig:alexresB}. 

\begin{enumerate}
    \item Account A has significantly more noise in its main feed as compared to account B, with most of these from suggested posts. We believe account A had more keywords that were of interest to FB. When account A searched for keywords like development, building and growth, FB showed pages that generally were more popular and posted a lot on their pages. 
    In essence, we believe that these specific interests likely bring FB more profit as they advertise exact products that a user can easily buy.
    \item Account B has significantly more noise related feed in its latest video feed. This, we believe is because account B had the keywords cooking, chef and kitchen which produced a lot of video feed about cooking recipes. 
\end{enumerate}

\noindent From these two observations, we can conclude that adding noise efficiently depends more on the keywords themselves than on the amount of likes and video watches. Nonetheless, with the acquired data, we proceeded to calculate the effective privacy. 
Posts from friends were ignored since those usually did not reflect any particular attribute of the user. By taking these results into account, the effective privacy yields a result of 0.06 for account A and 0.13 for account B. Both accounts seem to have achieved a larger degree of privacy, with account A basically having its noise attributes almost indistinguishable from its real attributes. This, according to \autoref{sec:QuantifyPrivacy}, means that account A has achieved privacy. Account B however, likely needs more time to run the tool. 



\smallskip
\noindent\textbf{Limitations:} 
After running experiments with \SHORT{}, we observed the following limitations in its usability. Firstly, the user's FB feeds become less appealing as they become more and more infested with noise related posts. The current solution for this would be to tone down the amount of noise the tool generates. As a future work, a noise filtering tool can be implemented as a browser extension to filter out the noise feed while using FB in the browser. 


Another limitation is the variety of noise data that a tool can generate automatically. In our implementation, we used liking posts, pages and watching videos as ways to generate noise. These were chosen as they represent direct interest of the topic at hand. Nevertheless, they are only a fraction of things that a normal user can do on FB. Other interactions such as user posting, commenting, reacting to posts and playing games are likely also used by FB for better user profiling. These interactions however are very hard to simulate adequately and can lead to unwanted issues such as generating posts that are unintelligible, inappropriate or even extremist, which, in the end, can lead to account blocking. 

It is also worth mentioning that some users may be hesitant to use \SHORT{} out of fear of liking something inappropriate. 
This is one of the reasons that the noise generated by the tool is not completely random. Mainly, the user chooses a seed keyword that will define the first noise attribute and start generating traffic based on it. The next noise attribute will be generated based on the initial seed -- a word that is related to the seed word. We admit that this might not be a perfect solution and solutions can be further developed in future works. Another concern is generating traffic related to illegal, extremist or abuse topics. This traffic however is constantly removed from the platform.

\section{Conclusion and Societal Impact}
\label{sec:Conclusion}
Social networks shaped the digital world becoming an indispensable part of our daily lives. Over the years, these platforms have gained a reputation for tracking user online activity. These strategies may prove threatening for multiple spheres of peoples' lives -- spanning from consumption to opinion formation -- and may have ominous effects on democracy~\cite{KHAN2021103112,9342970}. 
This vast collection of personal data by SNs is often exposed (i.e.\ sold) to third-party companies. 

In addition, SN users do not usually have a say on the information they access, as SNs prioritize the content presented on feeds, based on what users most probably want to see. In other words, SN algorithms seemingly hide content and have a great impact on the information users are able to reach. With privacy and societal concerns over SNs rapidly rising, these platforms are seen as rather controversial. 

Having identified these issues, we built \SHORT, a tool that adds new privacy safeguards for SN users aimed at hampering SN ability to serve targeted content. \SHORT{} allows users to define their desired level of privacy. 
In this way \SHORT{} strikes a balance between privacy and functionality. We believe this feature will be used in several services in the near future and will help towards building less biased SNs, while minimizing the amount of personal information processed by platforms.

\bibliographystyle{splncs04}
\bibliography{MetaPriv}

\end{document}